\documentclass{jfm}
\usepackage{graphicx}
\usepackage{epstopdf, epsfig}
\usepackage{xcolor}
\usepackage{amssymb}
\usepackage[normalem]{ulem}
\usepackage{hyperref}
\usepackage{tikz}
\usetikzlibrary{shapes}
\usetikzlibrary{positioning}
\newcommand{\yslant}{-0.6}
\newcommand{\xslant}{0.6}
\usepackage{caption}
\usepackage{color}

\shorttitle{Large-scale structures in high Reynolds number rotating Waleffe flow}
\shortauthor{S. Farooq, M. Huarte-Espinosa and R. Ostilla-M\'onico}

\title{Large-scale structures in high Reynolds number rotating Waleffe flow}

\author{ Shafqat Farooq\aff{1} \and  Martin Huarte-Espinosa\aff{2}
\and Rodolfo Ostilla-M\'onico\aff{1}
  \corresp{\email{rostilla@central.uh.edu}}}

\affiliation{\aff{1}Cullen College of Engineering, University of Houston, Houston, TX 77204, USA
\aff{2}University of Houston's Hewlett Packard Enterprise Data Science Institute, University of Houston, Houston, Texas 77204}

\begin{document}
\newcommand{\triangleup}{\raisebox{0pt}{\tikz{\node[draw,scale=0.3,regular polygon, regular polygon sides=3,fill=black,rotate=0](){};}}}
\newcommand{\triangledow}{\raisebox{0pt}{\tikz{\node[draw,scale=0.3,regular polygon, regular polygon sides=3,fill=blue,rotate=180](){};}}}
\newcommand{\squarefilled}{\raisebox{0.5pt}{\tikz{\node[draw,scale=0.4,regular polygon, regular polygon sides=4,fill=yellow](){};}}}
\newcommand{\circlefilled}{\raisebox{0.5pt}{\tikz{\node[draw,scale=0.4,circle,fill=green](){};}}}
\newcommand{\starfilled}{\raisebox{0.5pt}{\tikz{\node[draw,scale=0.4,star,fill=red](){};}}}

\maketitle

\begin{abstract}
We perform direct numerical simulations of rotating turbulent Waleffe flow, the flow between two parallel plates with a sinusoidal streamwise shear driving force, to study the formation of large-scale structures and the mechanisms for momentum transport. We simulate different cyclonic and anti-cyclonic rotations in the range of dimensionless rotation numbers (inverse Rossby numbers) $R_\Omega$ $\in$ $[-0.16, 2.21]$, and fix the Reynolds number to $Re=3.16\times 10^3$, large enough such that the shear transport is almost entirely due to Reynolds stresses and viscous transport is negligible. We find an optimum rotation in anti-cyclonic regime at $R_\Omega=0.63$, where a given streamwise momentum transport in the wall-normal direction is achieved with minimum mean energy of the streamwise flow. We link this optimal transport to the strength of large scale structures, as was done in plane Couette by Brauckmann \& Eckhardt (J. Fluid Mech., 815, 2017). Furthermore, we explore the large-scale structures and their behaviour under spanwise rotation, and find disorganized large structures at $R_\Omega =0$ but highly organized structures in the anti-cyclonic regime, similar to the rolls in rotating plane Couette and turbulent Taylor Couette flow. We compare the large scale structures of plane Couette flow and Waleffe flow, and observe that the streamwise vorticity is localized inside the cores of the rolls. We show that the rolls take energy from the mean flow at long time-scales, and relate these structures to eigenvalues of the streamfunction.
\end{abstract}


\section{Introduction}

While turbulent flows are generally chaotic and random, coherent large-scale motions can exist within them. Wall-bounded turbulent flows are not an exception, and large-scale, organized structures have been reported in both experiments and numerics \citep{jim12}. The study of structures was pioneered in channel flow, i.e.~the pressure driven flow between two parallel plates, a popular model to study wall-bounded turbulence. Using large-eddy simulation (LES), \citet{MOINKim82} found large-amplitude streamwise vortical structures concentrated near the wall. These large scale structures were attributed to a splatting effect (a net transfer of energy between perpendicular velocity fluctuations) and to Helmholtz-type instabilities of intense shear layers at and near the wall. Similar large-scale structures were also found by \citet{MOSMOIN84} using direct numerical simulation (DNS) in turbulent channel flow with a curved geometry, and by \citet{Kim87} in a fully turbulent channel. \citet{Kim87} further characterized these structures through local maxima and minima of the streamwise vorticity, i.e.~a streamwise vortex model. With the increased availability of computational power, the achievable Reynolds numbers and domain sizes in simulations have kept on growing. Large, coherent structures have still been observed even for the recent simulations at $Re_\tau  \approx 5200$ by \citet{Myoungku15}. Coherent flow structures have also been observed in other types of high Reynolds number wall-bounded flows, including pipe flows \citep{Bruno07}, plane Couette flow, i.e.~the shear flow between two parallel plates, in turbulent boundary layers \citep{smi11}, and in von Karman flow, i.e.~the flow between to coaxial rotating disks \citep{zandbergen1987karman,rav04}.

Taylor-Couette (TC) flow \citep{gro16}, the flow between two co-axial and independently rotating cylinders is another canonical wall-bounded flow where coherent large-scale structures are present. Their formation has usually been attributed to centrifugal (linear) instabilities since the seminal study by \cite{Taylor23}. Because of this, they are usually referred to as Taylor rolls. Due to the centrifugal effects, Taylor-Couette and its structures have usually been studied from the point of view of angular momentum convection, and not from the perspective of a wall-bounded flow \citep{Lathrop92}. A notable difference between Taylor rolls and more general structures in wall-bounded flows is that Taylor rolls are pinned, i.e.~they do not move around the fluid domain, and this is true with increasing Reynolds numbers up to the so-called turbulent Taylor rolls seen at $Re\simeq 10^6$ by \cite{hui14}. Turbulent Taylor rolls survive at high Reynolds number only for some combinations of curvature, and mild outer cylinder rotation \citep{OM14,hui14}. Remarkably, in a TC geometry with a large curvature, rolls do not exist at high Reynolds number for pure inner cylinder rotation \citep{OM14}. This means that something else aside from centrifugal effects must play a role. Numerical studies of TC flow conducted by \citet{Fran19} found that turbulent Taylor rolls appeared with a combination of shear and mild anti-cyclonic rotation. Their onset was not controlled by the curvature of the system. At high Reynolds numbers, Taylor rolls would unpin, or even disappear if anti-cyclonic rotation was removed. \cite{Fran19} also found that the rolls are persistent in the limit of vanishing curvature, i.e.~when Taylor-Couette flow becomes rotating plane Couette flow, if anti-cyclonic rotation is present. 

This showed that the study of turbulent Taylor rolls could be better approached from a shear flow perspective, and not simply by thinking of them as a continuation of the centrifugal linear instability seen at low Reynolds numbers. Indeed, low-curvature Taylor-Couette flow shows some characteristics of shear flows for Reynolds numbers just beyond the onset of the linear, centrifugal instability. Taylor rolls develop a streamwise modulation, after which they are usually denoted ``wavy'' Taylor vortices \cite{and86}. This is linked to appearance of large scale streaks \citep{des18}. Taylor rolls are then fed by the non-linear interaction of streaks. This non-linear interaction between the pinned Taylor roll and the streak was attributed to the activation of the self-sustained process (SSP) of shear flows \citep{des18}, which is described below. In this spirit, \cite{Fran19} showed that energy of turbulent Taylor rolls and streaks varied periodically with a distinct phase-shift, and a long-timescale multistage process energized the pinned structures. But despite the low Reynolds number link \citep{des18}, it is not clear how the high Reynolds number turbulent Taylor roll-streak process is related to the SSP responsible for the generation of turbulence in shear flows. 

The name SSP commonly refers to a multistage process responsible for regenerating wall-bounded turbulence, where streamwise rolls interact with streamwise velocity to cause streaks. These unstable streaks interact non-linearly, reinforcing the rolls and completing the SSP cycle. \cite{WALEFFE97} was the first to show that a generic process was responsible for the regeneration of turbulence in wall-bounded flows. Unlike earlier studies of the regeneration mechanisms in channel flow \citep{ham95}, \cite{WALEFFE97} studied the SSP in a fluid system where the flow is bounded by two infinite stress-free plates and forced using a body shear forcing force. \cite{WALEFFE97} was the first to study this system in detail to assess the role of the no-slip condition in the SSP, even if it had been used by \cite{tol36} to show that an inflection point was not a sufficient condition for linear instability \citep{dra04}. Because of this it has been recently associated to Waleffe's name \citep{beaume15,cha16} and we will refer to it as Waleffe flow from here on. 

In the spirit of \cite{WALEFFE97}, we set out to investigate whether the large-scale coherent structures of plane Couette and Taylor-Couette flow are part of a more general class of structures, which require only shear (and anticyclonic rotation), as the SSP does, or if they are something distinct, separated from the SSP because they require the presence of a no-slip wall. The natural system to investigate this is rotating Waleffe flow. 

The absence of a no-slip wall also provides for a second avenue of investigation. In TC and in rotating plane Couette (RPC) flow, the transport of torque or shear, is greatly enhanced by the presence of large-scale structures. In particular, in Taylor-Couette, the angular velocity current, non-dimensionalized as a Nusselt number ($Nu_\omega$), depends mainly on three parameters. First, the shear, which can be non-dimensionalized as a shear Reynolds number ($Re_s=U(r_o-r_i)/\nu$). Second, the solid-body system rotation, which appears in the equations as a Coriolis force, and its magnitude can be expressed non-dimensionally as a rotation number ($R_\Omega= 2\Omega (r_o-r_i)/U$). Finally, the curvature, expressed as a radius ratio $\eta=r_i/r_o$. Here, $r_i$ ($r_o$) is the inner (outer) cylinder radius, $U$ a characteristic shear velocity, $\nu$ the kinematic viscosity of the fluid and $\Omega$ the background rotation. Each of these parameters is linked to both the torque and to the presence of large-scale structures \citep{Van12,Brauckmann13,OM14}. In the low curvature regime ($\eta \geq 0.9$), where centrifugal forces are negligible, \citet{Brauckmann16} showed that at $Re_s\sim\mathcal{O}(10^4)$, there are two local maxima in the $Nu_\omega(R_\Omega)$ curve: one narrow and one broad. The ``broad maxima'' at $R_\Omega\approx 0.2$, dominated at lower shear Reynolds number ($Re_s$) and was related to the enhancement of large-scale vortical flow structures (\citet{Brauckmann16}). On the other side, the ``narrow peak'' at $R_\Omega=0.02$ was linked to a shear instability due to turbulent boundary layers (\citet{Brauckmann17}) and emerged with increasing $Re_s$. It was argued this narrow peak would supersede the broad peak at very high $Re_s$ (\citet{Brauckmann16, Brauckmann17}), and this was confirmed experimentally by \citet{Ezeta19}. 

The broad and narrow peaks in the shear/torque transport were found to exist even in the limit of rotating Plane Couette flow, when curvature was completely absent. If large-scale vortical structures similar to the so-called turbulent Taylor rolls appear in Waleffe flow, one could expect that in the regions of parameter space where they are strengthened, a similar shear transport enhancement will exist. And if it existed, this optimal transport would survive well into the turbulent regime as it would not be superseded by boundary layer instabilities from a no-slip wall. Thus the second and third questions we set out to answer are (i) does optimal transport exists in rotating Waleffe flow? (ii) If so, how is it linked to large-scale structure enhancement?

The paper is organized as follows. In $\S$2, we define the numerical set up, control parameters, spatial resolution and domain size study. These include details of the numerical scheme ($\S$2.1), energy spectrum studies to assess the spatial resolution ($\S$2.2) and autocorrelation studies ($\S$2.3) to assess the size of the domain. We then detail the results of our investigation in $\S$3, including a characterization of the transport of shear in $\S$3.1, the effect of rotation on the statistics of Waleffe flow at high Reynolds number in $\S$3.2 including a discussion of optimal transport,  the effect of rotation  on the large-scale structures, and how it is further linked to optimal transport and measures against plane Couette flow in $\S$3.3 and a further characterization of these structures in $\S$3.4. We conclude with a brief summary and an outline for further research.

\section{Numerical setup}

\subsection{Problem setup and non-dimensionalization}

We perform direct numerical simulations (DNS) of rotating Waleffe flow in a three dimensional domain which is bounded by free-slip walls in the $y$-direction at $y=0$ and $y=d$, and is periodic in the streamwise ($x$) and spanwise ($z$) directions with periodicity lengths $L_x$ and $L_z$ respectively. A body force \textbf{f} is used to force the flow. A Coriolis body force is used to simulate solid-body rotation in the flow, which can be either cyclonic, i.e. where the spanwise rotation vector is parallel to the vorticity of laminar base flow, ($R_\Omega <0$) or anti-cyclonic, i.e.~the spanwise rotation vector is anti-parallel to the vorticity of laminar base flow ($R_\Omega >0$).  With this, the Navier-Stokes equations thus read: 

\begin{equation}
 \frac{\partial \textbf{u}}{\partial t} + \textbf{u}\cdot \nabla \textbf{u}  + 2\Omega (\textbf{e}_z \times \textbf{u}) = -\nabla p + \nu \nabla^2 \textbf{u}+\textbf{f},
\end{equation}
 
\noindent which is solved alongside the incompressibility condition:
 
\begin{equation}
 \nabla \cdot \textbf{u} = 0,
\end{equation}

\noindent where $\textbf{u}$ is the velocity, $\Omega$ is the background spanwise rotation, $p$ the pressure and $t$ is time.  

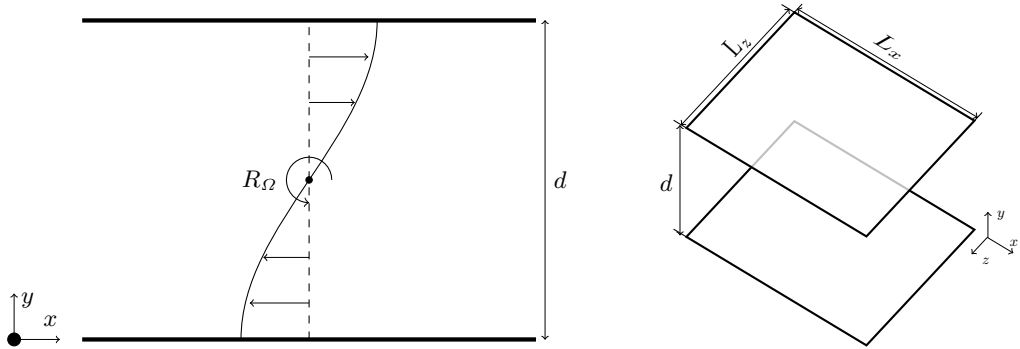
\begin{figure}
\begin{tikzpicture}[scale=0.6]
\draw[black,ultra thick](0,7)-- (10,7);
\draw[black,ultra thick](0,0)-- (10,0);


\draw[black,dashed](5,0)-- +(0,7);
\draw[rotate=90,shift={(0,-5)}] (0,1.5) cos (3.5,0) sin(7,-1.5);curve

\draw[->][black,ultra thin](5,0.8)--+ (-1.3,0);
\draw[->][black,ultra thin](5,1.8)--+ (-1,0);
\draw[->][black,ultra thin](5,5.2)--+ (1,0);
\draw[->][black,ultra thin](5,6.2)--+ (1.3,0);

\draw[->][black,domain=0:270] plot ({5+0.5*cos(\x)}, {3.5+0.5*sin(\x)});
\filldraw[black] (5,3.5) circle (2pt) node[anchor=east] {};

\filldraw[black] (-1.5,0) circle (4pt) {};
\draw[->][black,ultra thin](-1.5,0)--+ (0,1);
\draw[->][black,ultra thin](-1.5,0)--+ (1,0);
\draw[<->, ultra thin][black](10.2,0) to +(0,7);

		
\fill[black](-0.7,0.1) node [scale=1,anchor= south]{$x$};
\fill[black](-1.2,0.5) node [scale=1,anchor= south]{$y$};
\fill[black](10.2,3.5) node [scale=1,anchor=west]{$d$};
\fill[black](4.5,3.5) node [scale=1,anchor=east]{$R_\Omega$};
\end{tikzpicture}
\hspace{0.8cm}
\begin{tikzpicture}[scale=0.34]
\begin{scope}[ yshift=-120, every node/.append style={yslant=\yslant,xslant=\xslant},
		yslant=\yslant,xslant=\xslant
	] 
		\draw[black, thick] (0,0) rectangle (7,7);
		
        \draw[black](0,0) to (-0.5,0);
        
        \draw[->, ultra thin][black](7.5,7) to +(1,0);
        \draw[->, ultra thin][black](7.5,7) to +(0,-1); 
        
	\end{scope}
	
	
    \draw[<->][black,ultra thin](-0.25,0.1,0) to (-0.25,-4.15,0);
    
    \draw[->, ultra thin](7.49,-8.5,-11) to + (0,1,0);

	\fill[black](xyz cs:x=-0.2,y=-2.1,z=0) node [scale=1,anchor= east]{$d$};
	\fill[black](8,-8,-11) node [scale=0.6,anchor= south]{$y$};
	\fill[black](8.5,-9,-11) node [scale=0.6,anchor= south]{$x$};
	\fill[black](8.1,-9,-9.2) node [scale=0.6,anchor= south]{$z$};
	
	\begin{scope}[
		yshift=0,
		every node/.append style={yslant=\yslant,xslant=\xslant},
		yslant=\yslant,xslant=\xslant
	]
		\fill[white,fill opacity=.75] (0,0) rectangle (7,7); %
		\draw[<->,ultra thin][black](-0.2,0) to (-0.2,7);
		\draw[black](0,0) to (-0.5,0);
		\draw[black](0,7) to + (-0.5,0);
		
		\draw[<->,ultra thin][black](0,7.2) to (7,7.2);
		\draw[black](0,7) to + (0,0.5);
		\draw[black](7,7) to + (0,0.5);
		Opacity
		\draw[black, thick] (0,0) rectangle (7,7); 

		
		\fill[black](-0.7,3.5) node [rotate=90, scale=1,anchor= west]{$L_z$};
		\fill[black](3.2,7.2) node [ scale=1,anchor= south]{$L_x$};
		
	\end{scope} 
\end{tikzpicture}

\caption{Left: Two-dimensional spanwise cut of the system showing the streamwise shear force and background rotation of the Waleffe flow system. The $y$-origin of the system is shown as a thick black circle. Right: Three-dimensional view of the simulation geometry. }
\label{fi:Geom}
\end{figure}

The geometrical configuration and the input body force of Waleffe flow is show in figure \ref{fi:Geom}. The velocities in the $x$, $y$, and $z$ directions are denoted by $u$, $v$ and $w$ respectively. 
A streamwise shear body force is required to force the flow as, unlike plane Couette, flow no energy is injected through the walls. A sinusoidal profile is chosen, i.e.~$\textbf{f}= F \cos (\beta y) \textbf{e}_x$, with $\beta =\pi/d$, analogous to the setup used in \citet{WALEFFE97}. This means the force is maximum, but in opposite directions at both slip walls, and zero at the mid-plane. The force must be zero-average as otherwise the flow would constantly accelerate unopposed by the free-slip walls. 

Furthermore, in the absence of the Dirichlet (no-slip) boundary condition, the system is completely Galilean invariant. An arbitrary translation velocity can be added in either the streamwise or spanwise direction with no effect. A reference frame must be chosen, and we take the one for which the mean streamwise and spanwise velocities are zero. We will denote with the operator $\langle ... \rangle$ a streamwise, spanwise and temporal average. In this spirit, $\langle \phi \rangle$ represents a mean quantity, $\phi^\prime$ the instantaneous fluctuation around the mean quantity, and $\langle \phi^\prime \rangle$ the root-mean-squared fluctuation around this mean.

The equations are non-dimensionalized using the distance between the walls $d$ and the forcing amplitude $F$. We define a characteristic velocity $\tilde{U}$ for non-dimensionalization as $\tilde{U}=\sqrt{Fd}$. We note that we use a different characteristic velocity than the one classically used for Waleffe flow (c.f.~\citet{beaume15}), as we focus on the fully turbulent case, and not in perturbations around the laminar state. This definition results in a Reynolds number defined as $Re = \tilde{U}d/\nu = \sqrt{Fd^3}/\nu$ which is our first control parameter. The second non-dimensional control parameter accounts for the cyclonic and anti-cyclonic rotation of the system, and is defined as $R_\Omega = 2\Omega d/\tilde{U} =\Omega d/\sqrt{Fd}$. For this study, we fix $Re=3.16\times10^3$, large enough such that the flow is fully turbulent and that the shear transport takes place purely through Reynolds stresses (cf. $\S$3.1), and vary $R_\Omega$ in the range $[-0.16,2.21]$ to study the effect of rotation. After the initial transient, we run the simulations  between $200$ and $250$ $d/\tilde{U}$ time units to collect statistics.

The equations are discretizated in space using a second-order energy-conserving centered finite difference scheme, while temporal discretization is done using a third-order Runge-Kutta for the explicit terms and a second-order Adams-Bashforth scheme for the implicit viscous terms in the wall-normal direction. The simulation code used is based on the highly parallel FORTRAN-based AFiD (\url{www.afid.eu}) which has being used mainly for simulating turbulent Rayleigh-B\'enard convection and Taylor-Couette flow \citep{poe15}. This code has being comprehensively validated. Detailed information regarding the code algorithms can be found in \citet{ver96,poe15}. 

\subsection{Resolution study}

For determining what is an adequate spatial resolution of the flow, a series of simulations were performed at $Re = 3.16 \times 10^3$, for both no rotation ($R_\Omega=0$) and mild anti-cyclonic rotation ($R_\Omega = 0.32$). We can expect the wall-normal resolution to be less stringent in Waleffe flow due to the absence of the no-slip boundary condition. However, the streamwise and spanwise directions were found to be more restrictive than a plane Couette flow simulation at comparable Reynolds numbers. Indeed if one compares the Kolmogorov length-scale $\eta_K$ at $Re\sim 3\times 10^3$ for both systems, we obtain that $\eta_K$ is between five and six  times smaller for non-rotating Waleffe flow than for non-rotating plane Couette flow.

\begin{figure}
 \includegraphics[trim=0cm 20 0cm 0,clip,width=0.49\textwidth]{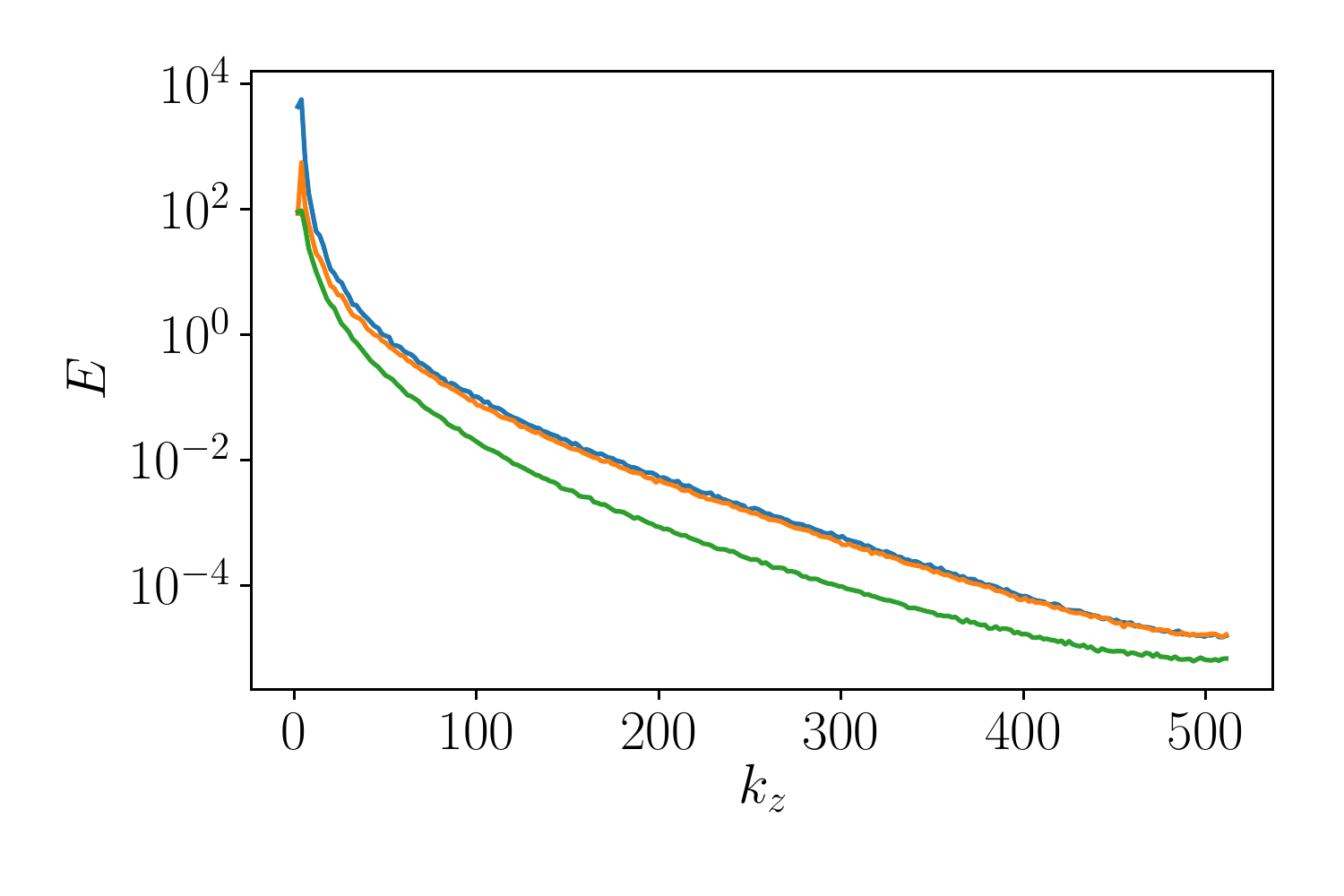}%
 \includegraphics[trim=0cm 20 0cm 0,clip,width=0.49\textwidth]{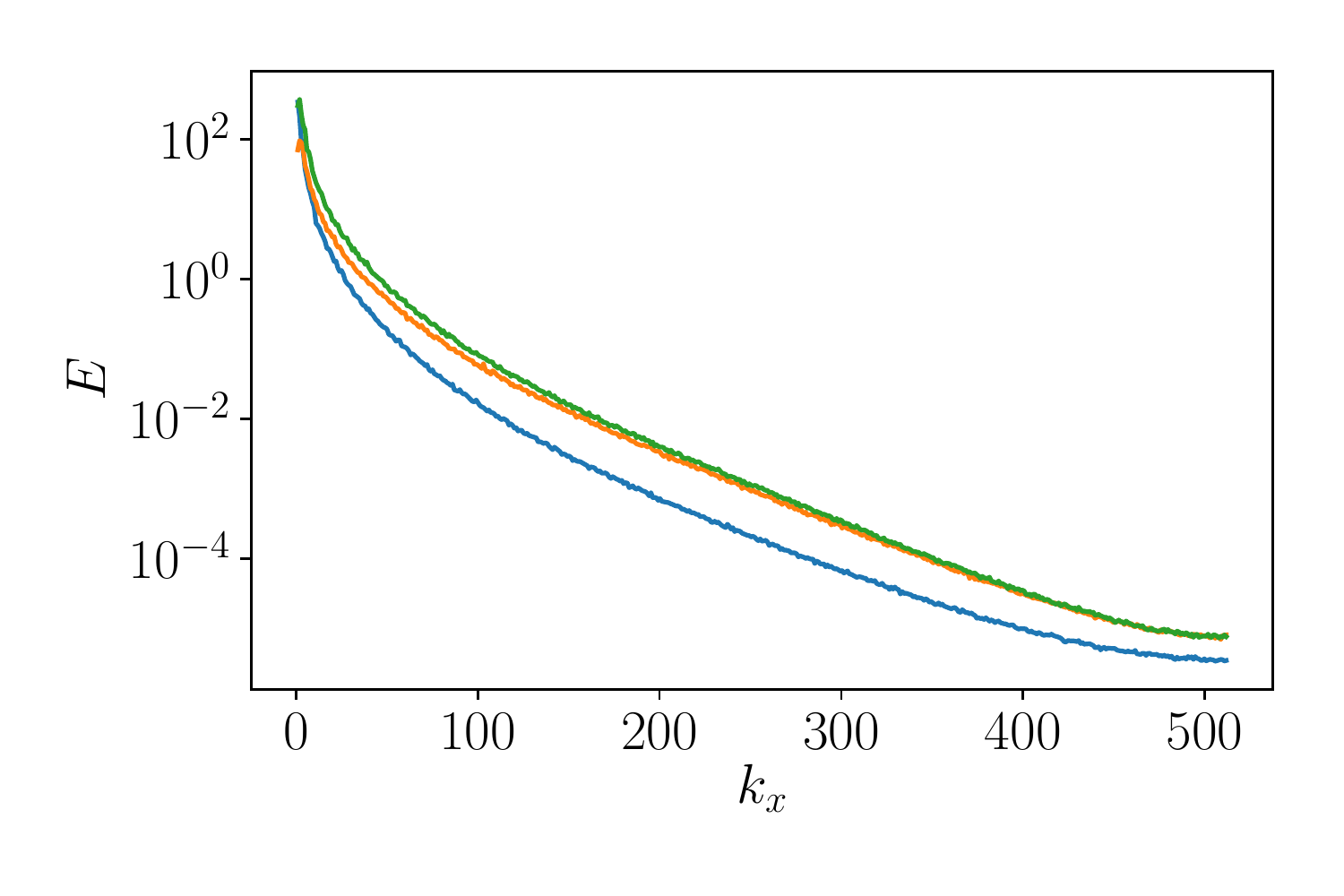}
 \caption{Energy spectra for the streamwise velocity $u$ (blue), the wall-normal velocity $v$ (orange) and the spanwise velocity $w$ (green) in the spanwise (left) and streamwise (right) directions at the mid-gap at $R_\Omega = 0$.}
\label{fig:Resol_Ro0} 
\end{figure}

Adequate resolution for the streamwise and spanwise directions was ensured through a spectral analysis of velocity data. We found that for $L_x/d=2\pi$ and $L_z/d=\pi$, $N_x=1024$ and $N_z=512$ points distributed uniformly were enough to accurately represent the velocity spectra at mid-gap at both no rotation, and $R_\Omega=0.32$. An extended dissipative regime at high wavenumbers, with energy $E\sim \exp(-k)$, is seen in Figure \ref{fig:Resol_Ro0} for both the streamwise and the spanwise directions. 

We note that both homogeneous directions have the same effective grid-spacing, $\Delta/d= 6.13\times 10^{-3}$. Non-dimensionalized by the Kolmogorov length-scale, this is around $\Delta/\eta_K \approx 2.51$. Using this grid spacing in the wall-normal direction would result in a grid of $N_y\approx 163$ points. For safety, we use $N_y=384$, and cluster points near the wall, such that the minimum wall-normal grid spacing is $\Delta/d=3.84\times10^{-4}$ and the maximum wall-normal grid spacing is $\Delta/d=3.85\times10^{-3}$, or $0.157 \lesssim \Delta/\eta_K \lesssim 1.58$ in Kolmogorov units.

\subsection{Domain periodicity study}

We performed a domain-size study in order to quantify the effect of the spanwise and streamwise periodicity lengths on the flow field statistics and on the large structures which might be present in the flow. We simulated several domains sizes, where $L_x$ and $L_z$ were doubled each time to produce larger and larger domains. The spatial resolution in both $x$ and $z$ directions was also doubled every time the domain was doubled, to keep the resolution from the previous paragraph. We refer to the domains henceforth as very small ($L_x/d=\pi$ and $L_z/d=\pi/2$), small ($L_x/d=2\pi$ and $L_z/d=\pi$), medium ($L_x/d=4\pi$ and $L_z/d=2\pi$) and large ($L_x/d=8\pi$ and $L_z/d=4\pi$).  We also note that the run time required to obtain adequate statistics does not decrease with domain size. The evolution of large-scales takes place in long time-scales (cf.~$\S$\ref{sec:tevolsc} for more details), and this strongly affects the value obtained for the mean streamwise velocity.

\begin{figure}

\includegraphics[trim=0cm 26 0cm 0, clip,width=0.5\textwidth]{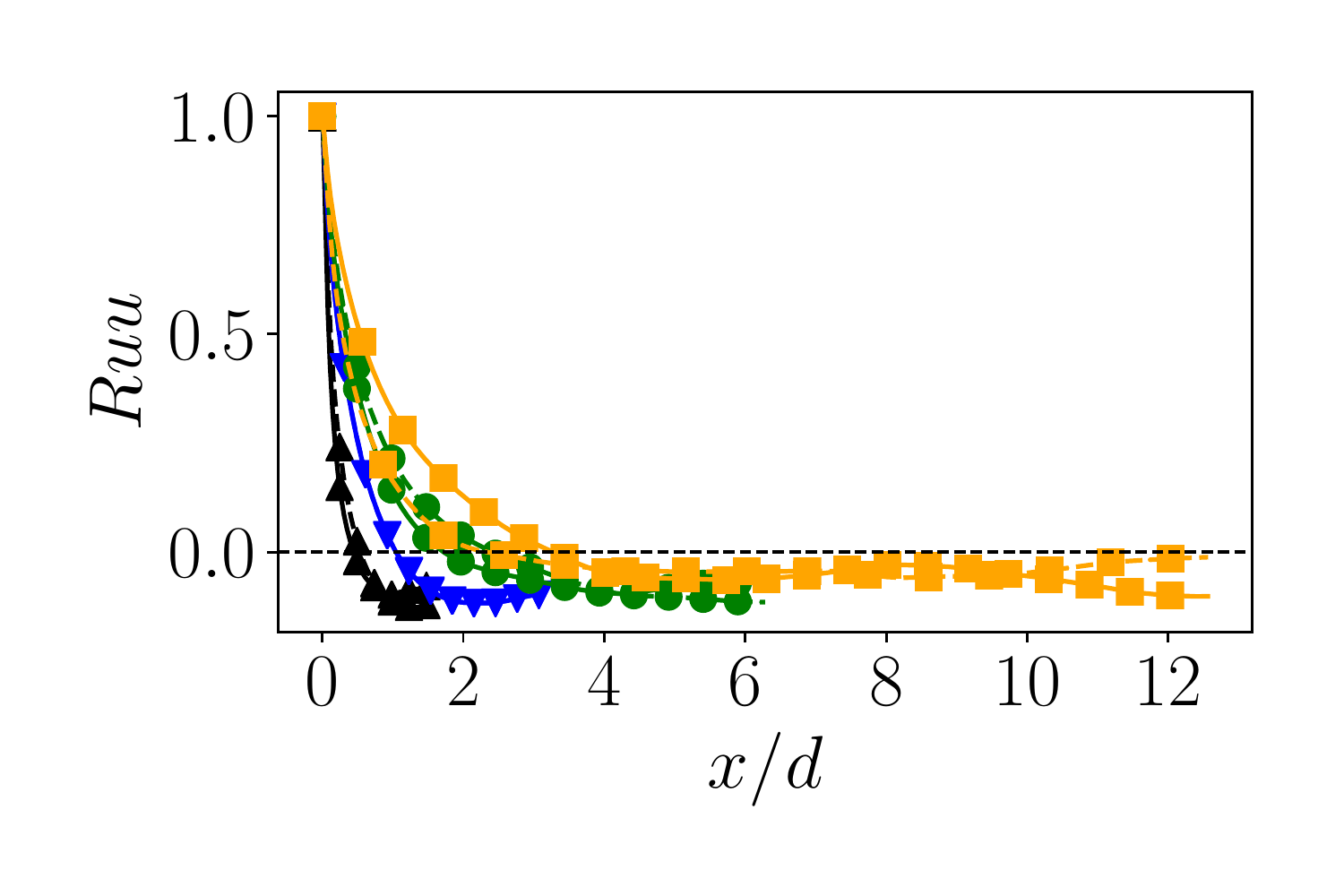}%
\includegraphics[trim=0cm 26 0cm 0, clip,width=0.5\textwidth]{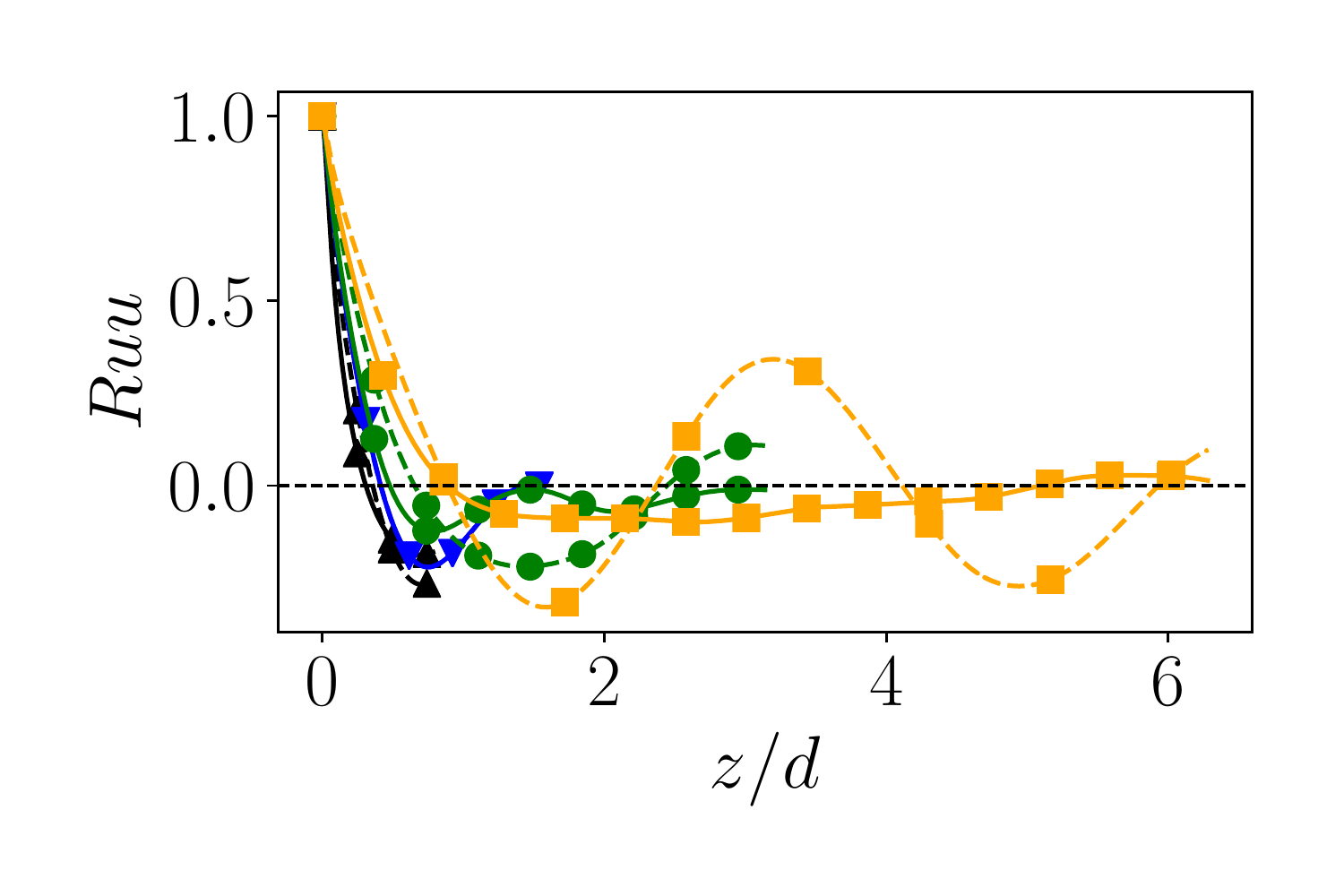}%
\\
\includegraphics[trim=0cm 26 0cm 0,
clip,width=0.5\textwidth]{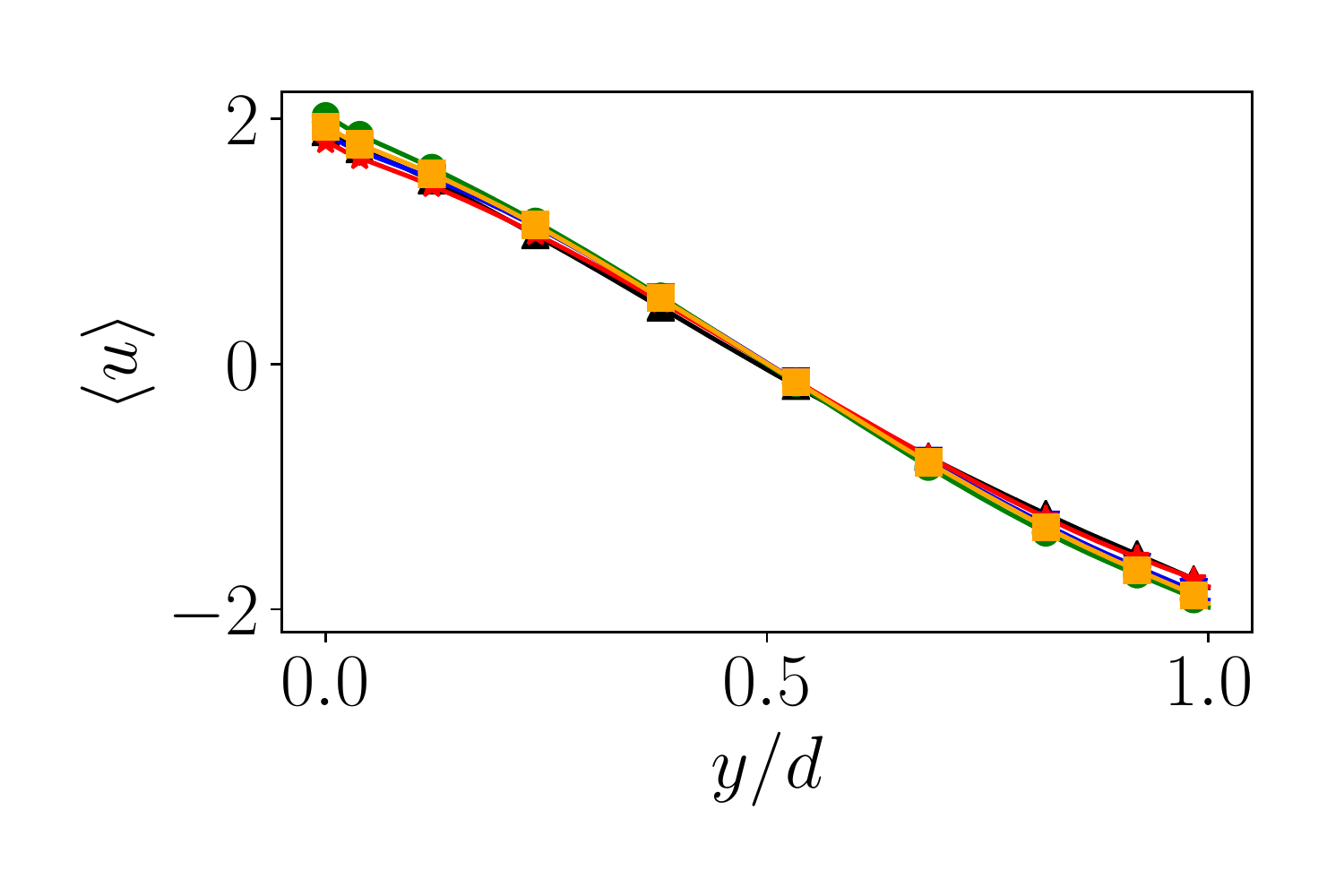}%
\includegraphics[trim=0cm 26 0cm 0, clip,width=0.5\textwidth]{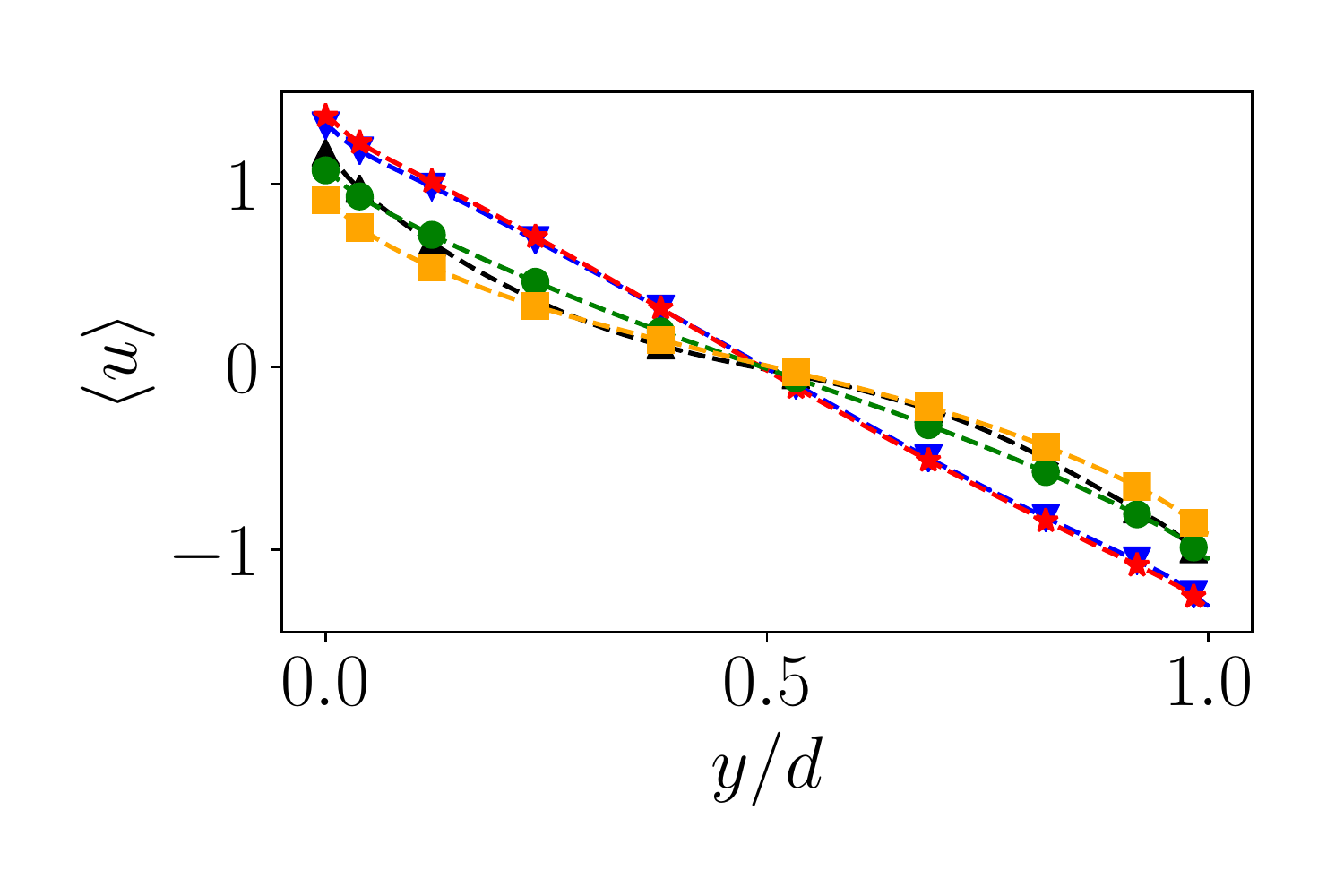}

\caption{The top panels show the streamwise velocity autocorrelation in the streamwise (left) and spanwise (right) directions. Solid lines are without rotation ($R_\Omega=0$) while dashed lines are with mild rotation ($R_\Omega=0.32$). Black upper triangle (\protect\triangleup): $L_x/d=\pi$, $L_z/d=\pi/2$, blue lower triangle (\protect\triangledow): $L_x/d=2\pi$, $L_z/d=\pi$, green circle (\protect\circlefilled): $L_x/d=4\pi$, $L_z/d=2\pi$;  yellow square (\protect\squarefilled): $L_x/d=8\pi$, $L_z/d=4\pi$. The bottom panel show the magnitude of averaged streamwise velocity for non-rotating (left) and $R_\Omega=0.32$ (right). The three-halves domain (\protect\starfilled): $L_x/d=3\pi$, $L_z/d=\frac{3}{2}\pi$ is also included.} 
\label{Aut:Waleffe} 
\end{figure}

The effect of the domain size on the results was checked in several ways. First, the top panels figure \ref{Aut:Waleffe} show the streamwise velocity autocorrelation in the streamwise and spanwise directions. The behaviour of non-rotating Waleffe flow is quite similar to what is commonly seen for plane Couette flow in the streamwise direction, with long decorrelation wavelengths. A strong effect of both rotation and domain size is seen in the autocorrelations, showing that the domain size will affect the behaviour of the structures inside the flow, and that rotation has a crucial effect on large-scale structures. All domains are sufficiently long in both the streamwise and spanwise dimensions for the velocity autocorrelations to change sign at least once. However, the domains are not large enough to show full decorrelation.

Another way to check domain-size independence is done by simply comparing the mean velocities obtained from the different computational domains. Due to the presence of large-scale structures which fill up the domain, the velocity profiles could be affected by the wavelength of these structures. Doubling the domain would not account for the changing wavelength of these structures, as the domain would simply be filled up with twice as many structures with the same wavelength. To avoid this, we run an additional case with periodicity lengths ($L_x/d=3\pi$ and $L_z/d=\frac{3}{2}\pi$), denoted as the three-halves domain.

In the bottom panels of \ref{Aut:Waleffe}, we show the streamwise, spanwise and temporally averaged streamwise velocity $\langle u \rangle$ for all cases. Remarkably, for the non-rotating case, the average streamwise velocity for all domains from the smallest domain to the largest collapse on top of each other. For the rotating case, there is collapse between the small and three-halves domain, while the other domains show strong variability.

\begin{figure}
\begin{center}
\includegraphics[trim=1cm 35 0cm 40, clip, height=0.15\textwidth]{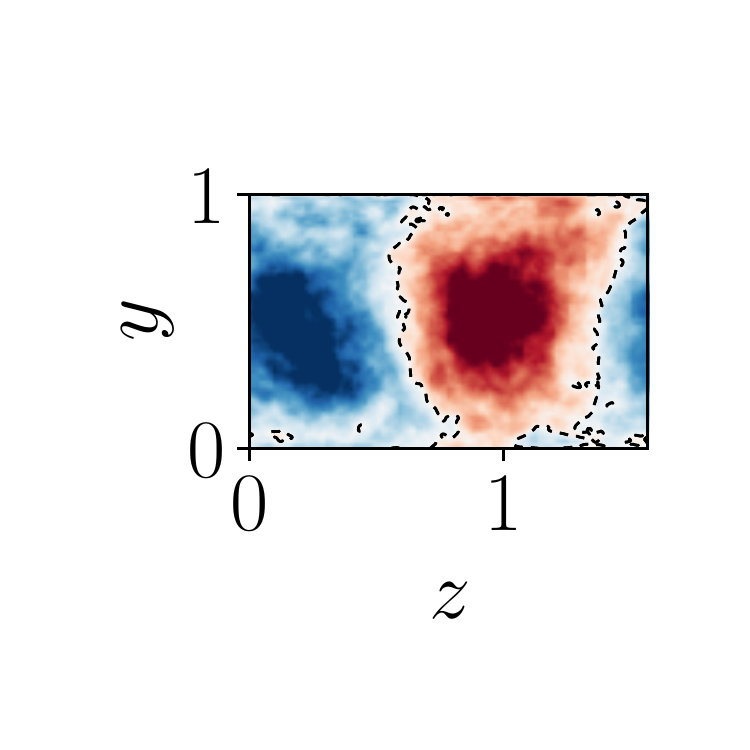}%
\includegraphics[trim=1cm 28 0cm 50, clip,height=0.18\textwidth]{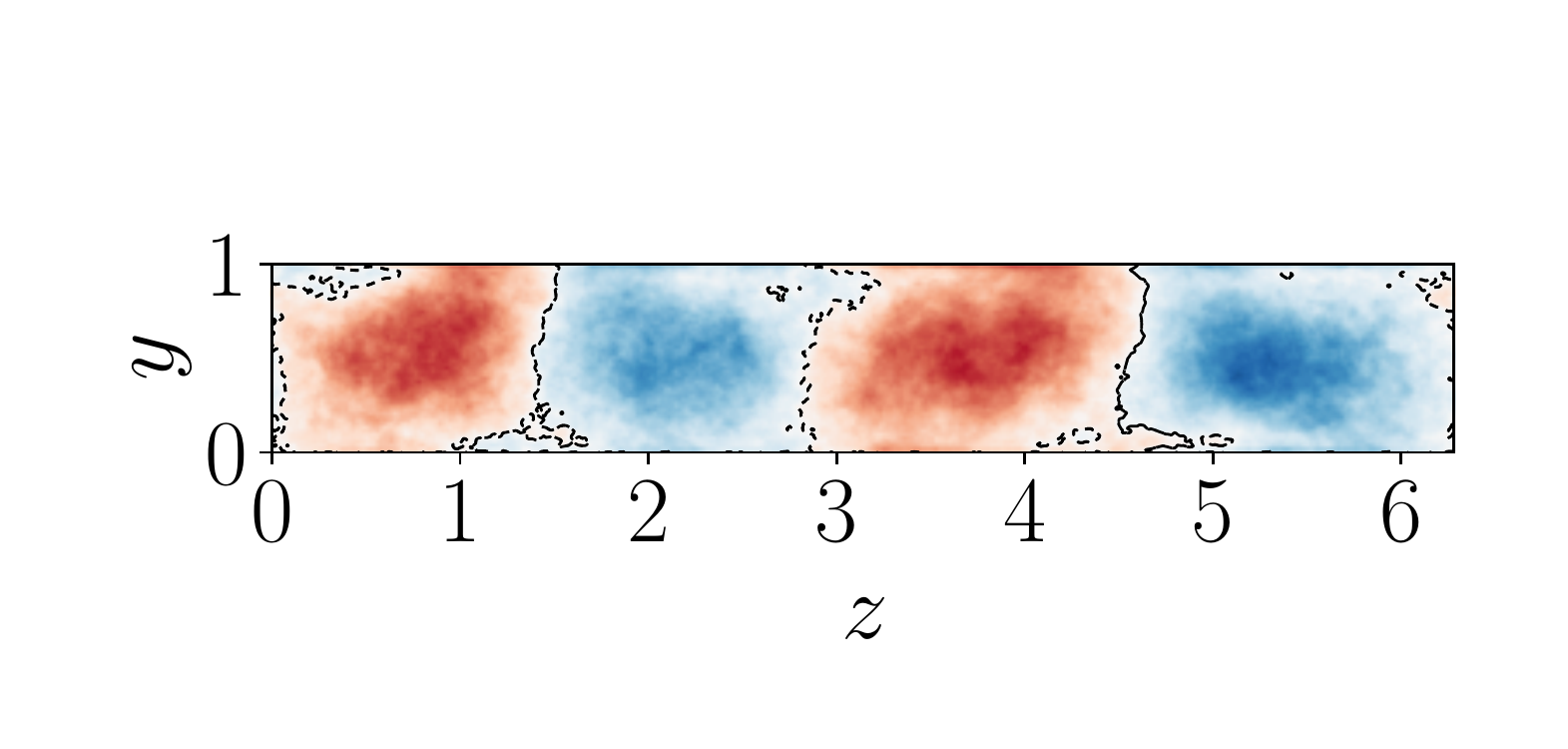}\\
\includegraphics[trim=1cm 28 0cm 40, clip, height=0.15\textwidth]{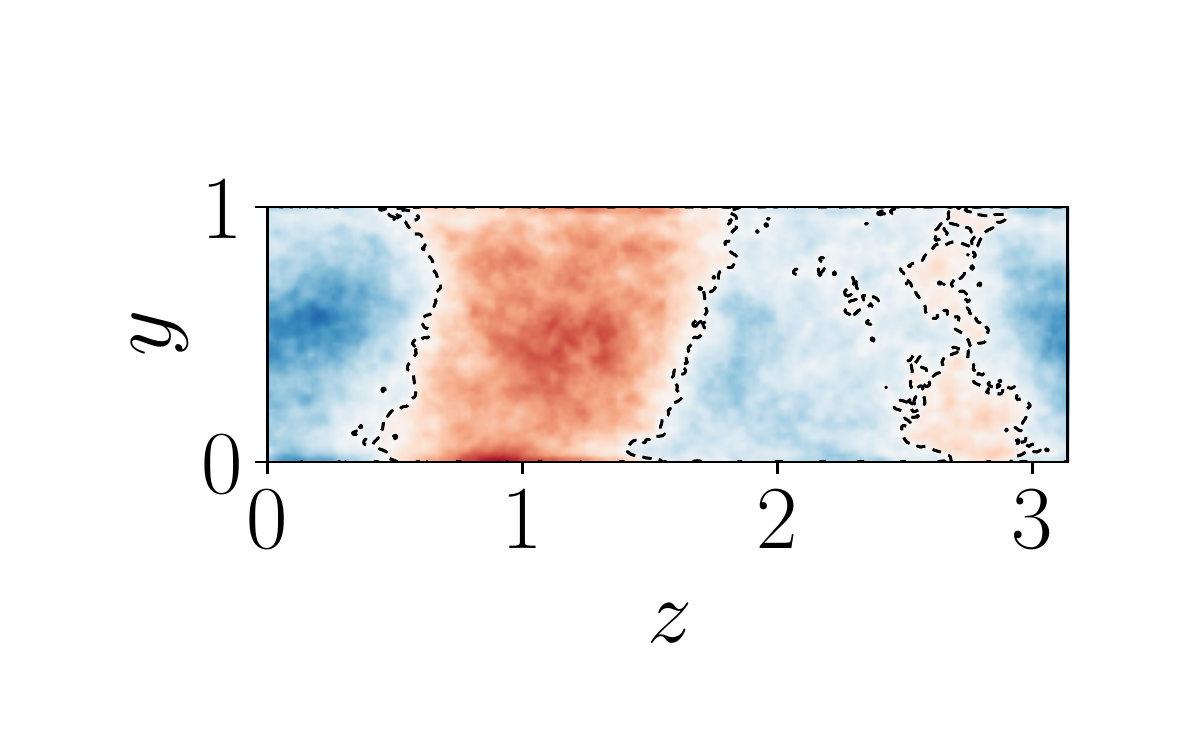}%
\includegraphics[trim=1cm 28 0cm 40, clip, height=0.16\textwidth]{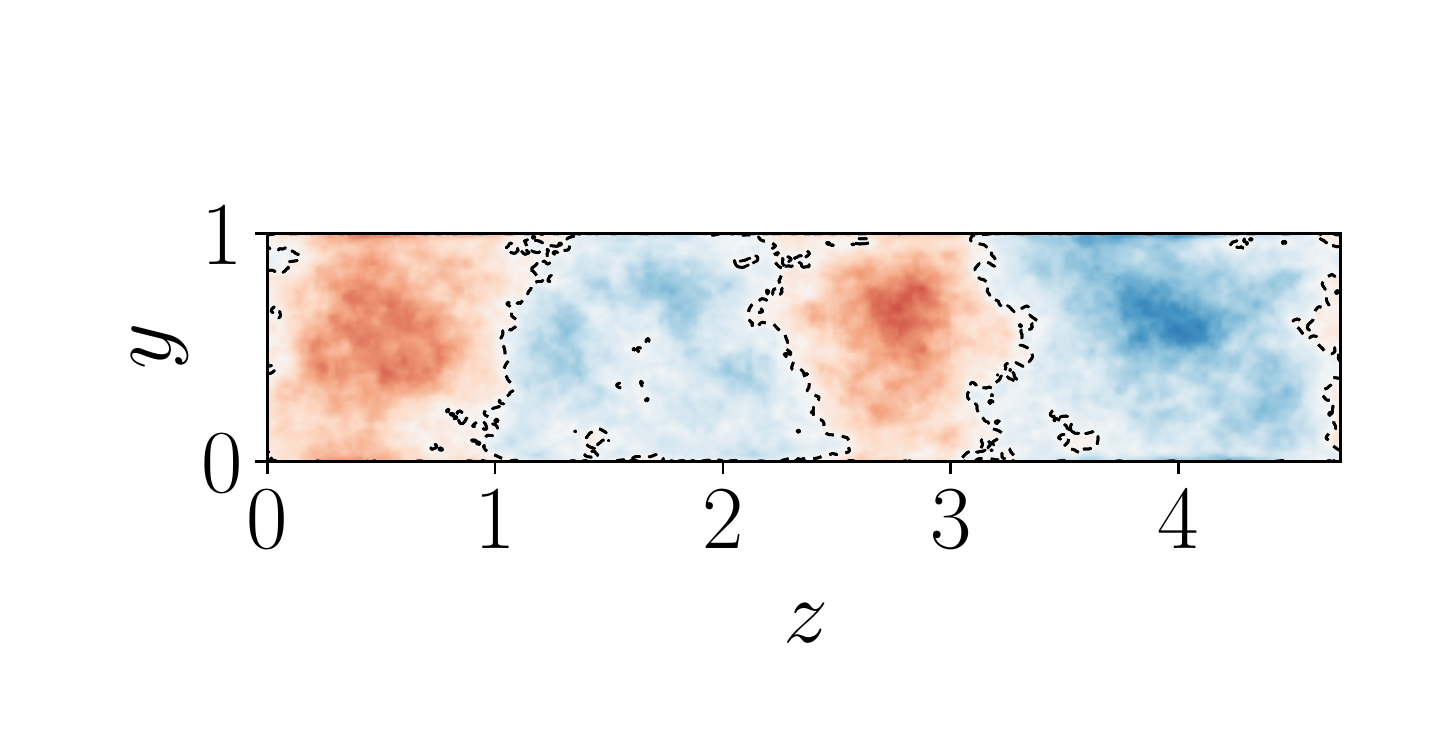} \\
\includegraphics[trim=4cm 10 3cm 20, clip, width=0.99\textwidth]{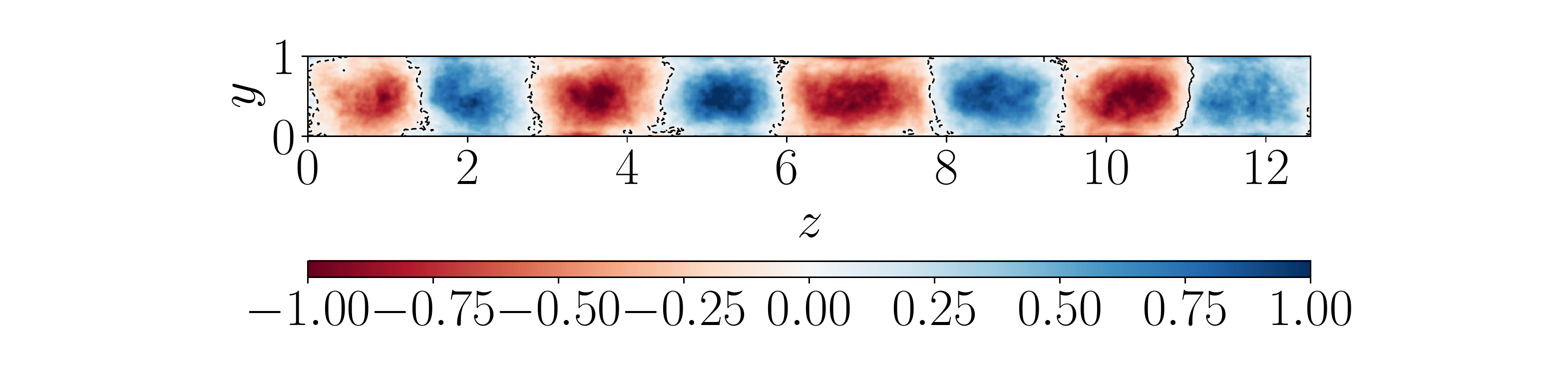}%
\end{center}
\caption{Pseudocolor plot of $\Omega_x$ for the five domain sizes simulated, at $R_\Omega=0.32$. Contours levels for vorticity are shown at zero to highlight the large-scale structures.}
\label{fi:Domain} 
\end{figure}

To understand this variation between the domains which only appears for the rotating case, we analyze the domain-size effects on the strength of large-scale pinned structures. In figure \ref{fi:Domain}, we show visualizations of the temporally and streamwise averaged streamwise vorticity $\Omega_x$. Why this definition of $\Omega_x$ captures the large-scale structures is discussed more elaborately in $\S$\ref{sec:rotlsc} but for now we note that it highlights streamwise-invariant structures which are pinned in the spanwise direction. 
A strong pinned structure is prominent for the very small domain, due to the constrained periodicity. The structure is somewhat weakened for the small domain, and the colors are less intense on the figure. The three-halves domain shows an organized pinned structure at three-halves the wavelength of the small domain, which remarkably does not affect the mean flow statistics of Figure \ref{Aut:Waleffe}. The medium domain and large domains present a larger number of organized structures, consistent with what is seen in the autocorrelations. Even if the medium and small domain have structures with the same wavenumber, the number of structures is different. This indicates that what is causing the differences in velocities must not be within the structures themselves, but in their interactions with each other, and the near-wall regions. 

We will not be able to get completely rid of domain-size effects as this study shows. From here, we proceed with the ``small'' domain with a single structure, as it allows us to explore a large parameter space while running the simulations for long times to gather enough statistics. We acknowledge that domain-size effects are unavoidable.

\section{Results}
\subsection{The nature of shear transport}

First, we show that the shear force in large Reynolds number Waleffe flow is transported almost exclusively by Reynolds stresses. In the statistically stationary regime, the mean velocities do not depend on time. We can write down an average of the total shear $\tau$ transported across a wall-normal plane, which is:

\begin{equation}
 \langle \tau \rangle = \mu \frac{d\langle u \rangle}{dy} + \rho \langle uv \rangle,
 \label{eq:tau}
\end{equation}

\noindent where $\rho$ is the fluid density and $\mu$ is the dynamic viscosity. Equation \ref{eq:tau} just states that in the statistically stationary regime, shear is transported through either viscosity or through Reynolds stresses. 

In high Reynolds number Waleffe flow, we may assume that the viscous shear transport ($\mu d\langle u \rangle/dy$) is negligible. This is because as $Re$ increases, the velocity gradients remain approximately constant. No viscous boundary layer where the average velocity has a sharp gradient is formed, because there is no wall with a no-slip condition. Hence, in the turbulent regime, the magnitude of the viscous term in Equation \ref{eq:tau} is $\mathcal{O}(Re^{-1})$ smaller than that of the Reynolds stress term. With this, Equation \ref{eq:tau} becomes: 

\begin{equation}
 \langle \tau \rangle \approx \rho \langle uv \rangle.
\end{equation}

By differentiating Equation \ref{eq:tau} with respect to the wall-normal direction, and comparing it to the Navier-Stokes equations, we have that the shear transported must be balanced by the body force:

\begin{equation}
 \frac{d \langle \tau \rangle}{dy} \approx \frac{d (\rho \langle uv \rangle) }{dy} \approx \rho \textbf{f},
\end{equation}

\noindent and solving the above equations gives an analytic expression for the Reynolds stress:

\begin{equation}
 \langle uv \rangle =\frac{F}{\beta}\sin(\beta y).
 \label{eq:theouv}
\end{equation}

\begin{figure}
\includegraphics[trim=0cm 0 0cm 0, clip,width=0.4\textwidth]{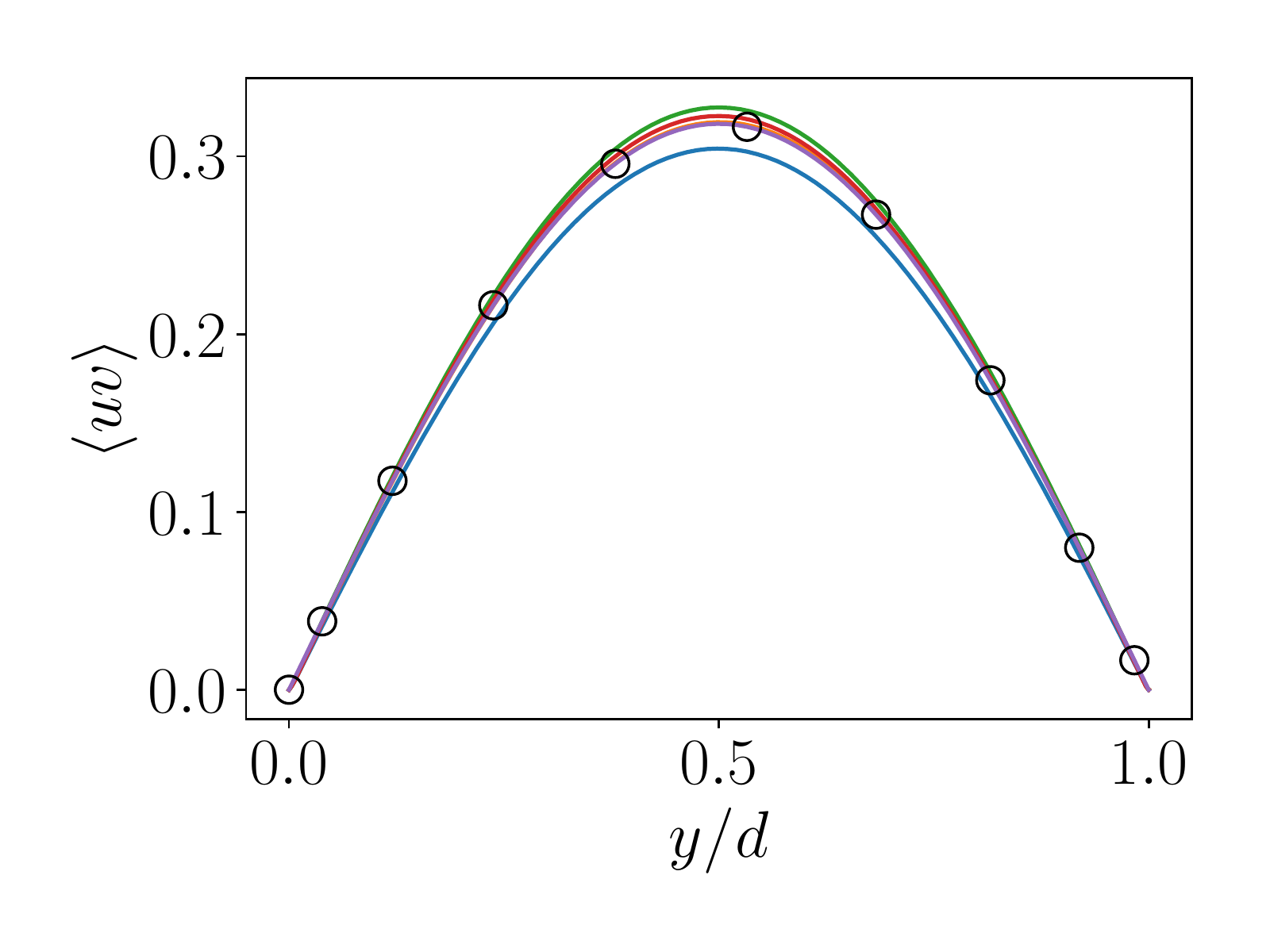}%
\includegraphics[trim=10cm 60 0cm 0, clip,width=0.185\textwidth]{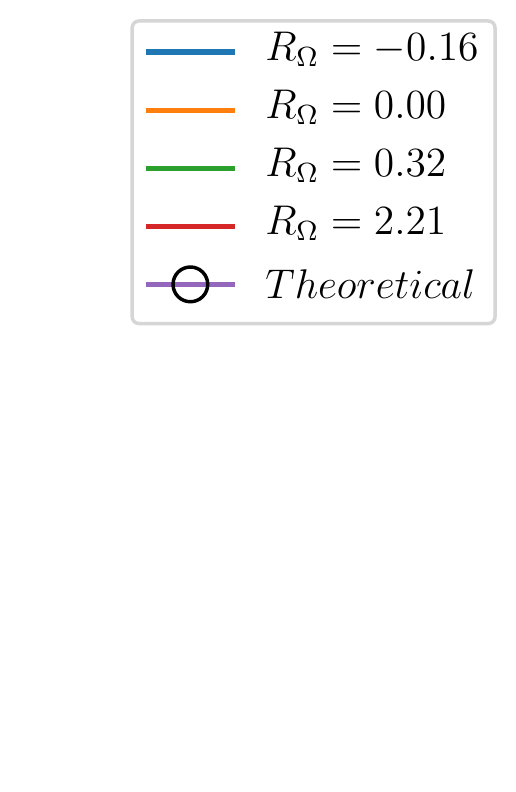}%
\includegraphics[trim=0cm 0 0cm 0, clip,width=0.4\textwidth]{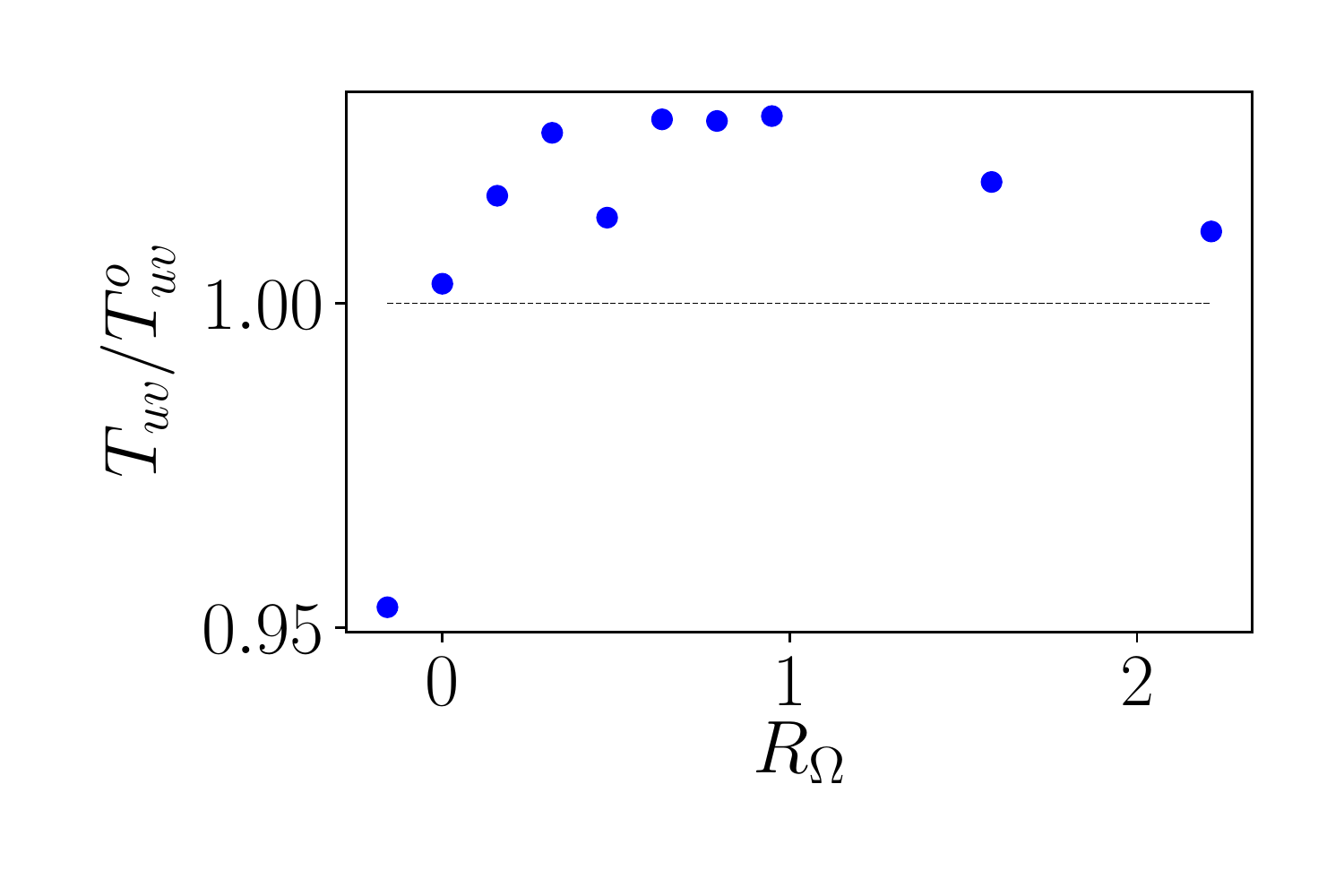}%

\caption{ Left: Averaged transport for different $R_\Omega$ values. The theoretical value for fully turbulent shear transport is shown with hollow circles. Right: Normalized integrated transport $T_{uv}$ for different $R_\Omega$ values. }
\label{fi:AvgVelMom} 
\end{figure}

This is valid in the statistically stationary regime only if our assumption that shear transport is fully due to Reynolds stresses. We check this in the left panel of Figure \ref{fi:AvgVelMom}, where we show the $\langle uv \rangle$ Reynolds stress components for different rotation numbers, as well as the theoretical value for $\langle uv \rangle$ from Equation \ref{eq:theouv}. We find that $\langle uv \rangle$ is almost equal to the theoretical value for full shear transport due to Reynolds stresses for all rotation numbers shown, even if some deviations exist for $R_\Omega=-0.16$, i.e.~cyclonic rotation. 

To further quantify transport, we define $T_{uv}$, as the integrated momentum transport in the wall-normal direction:

\begin{equation}
 T_{uv} = \displaystyle\int_0^d \langle uv \rangle~dy,
\end{equation}

\noindent and calculate its deviation from the analytic value for purely turbulent transport $T^o_{uv}=2F/\beta^2\approx 0.202Fd^2$. We show this quantity in the right panel of Figure \ref{fi:AvgVelMom}. The numerical value of $T_{uv}$ are approximately within $3\%$ of the theoretical value for different rotation numbers, except for $R_\Omega=-0.16$, corresponding to the case with cyclonic rotation. This tells us two things: First, that as seen in plane Couette and Taylor Couette, cyclonic rotation hampers turbulence and in this case, the viscous transport accounts for $\sim 5\%$ of the total transport. Second, that for no rotation or anti-cyclonic rotation, the shear transport is fully turbulent and $T_{uv}$ instead gives us an estimate for the temporal convergence errors in the simulations, as $T_{uv}/T^o_{uv}$ is close to unity. From the right panel of Figure 5, these can be estimated at around 2-3\%.

Unlike previous studies of rotating plane Couette flow \citep{Brauckmann16}, where the transported shear was a response of the system, an optimum momentum transport cannot be deduced from $T_{uv}$, because this is an input of the simulation, and $T_{uv}\approx T_{uv}^o$ in all cases. To define an optimum momentum transport, we must turn towards other diagnostics. This is further investigated in $\S$3.2 and $\S$3.3.

\subsection{Velocity magnitudes and optimum shear transport}

\begin{figure}
 \includegraphics[trim=0cm 0 0cm 0, clip,width=0.4\textwidth]{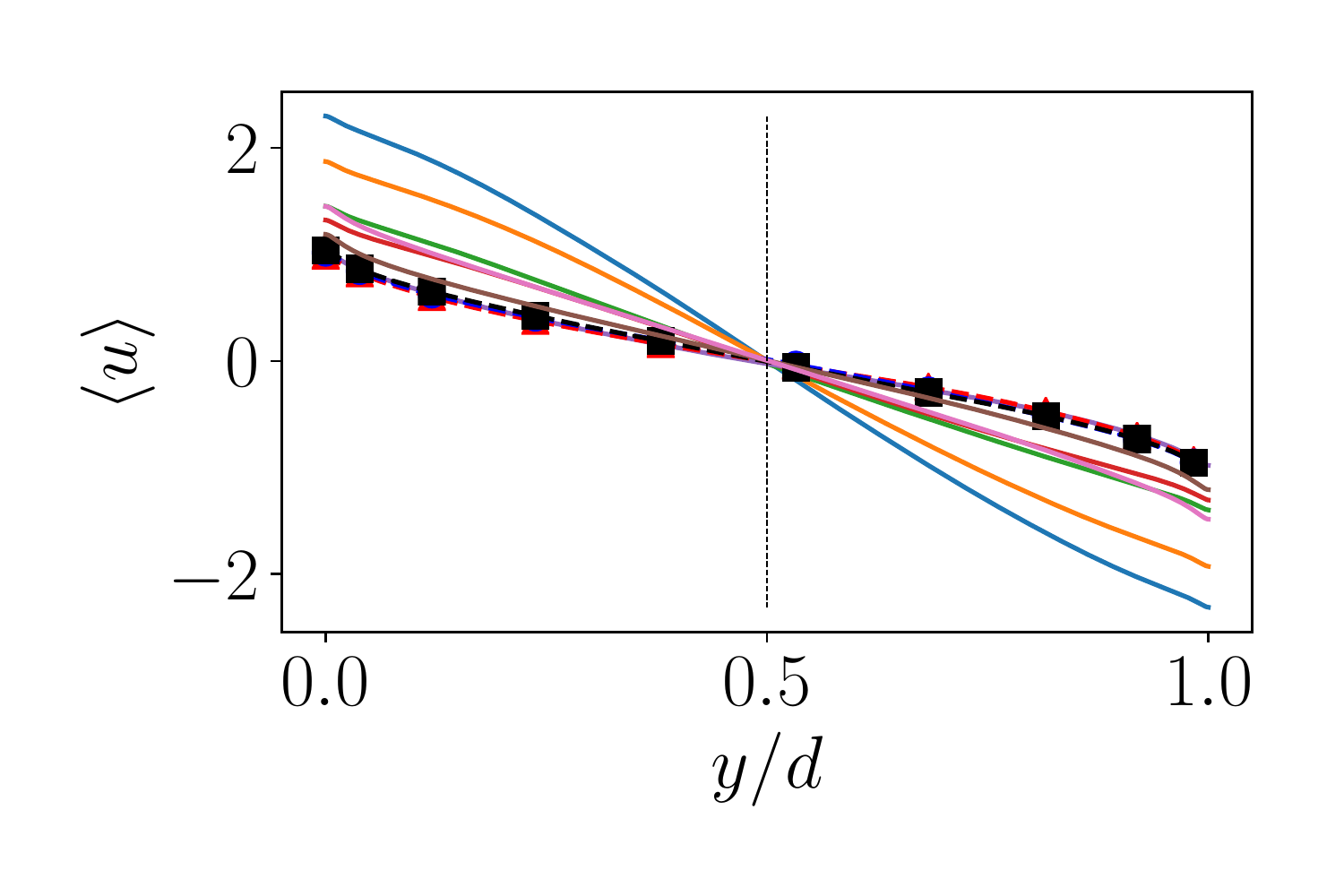}%
 \includegraphics[trim=10cm 60 0cm 0, clip,width=0.16\textwidth]{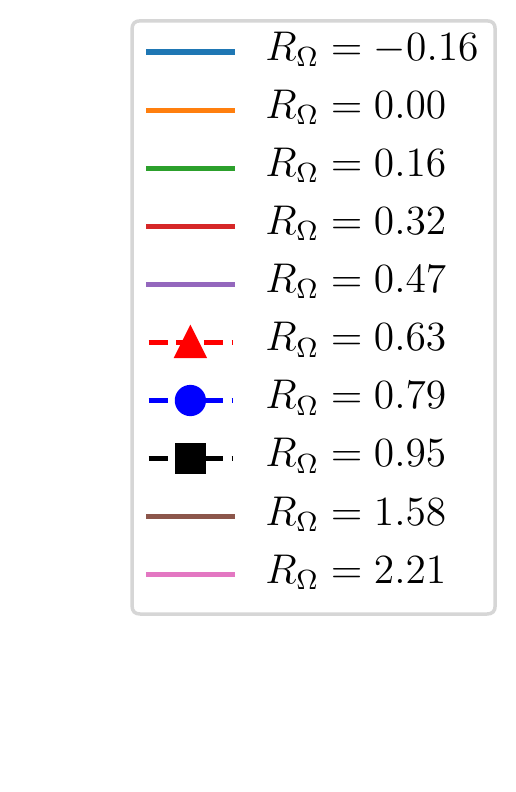}%
 \includegraphics[trim=0cm 0 0cm 0,clip,width=0.4\textwidth]{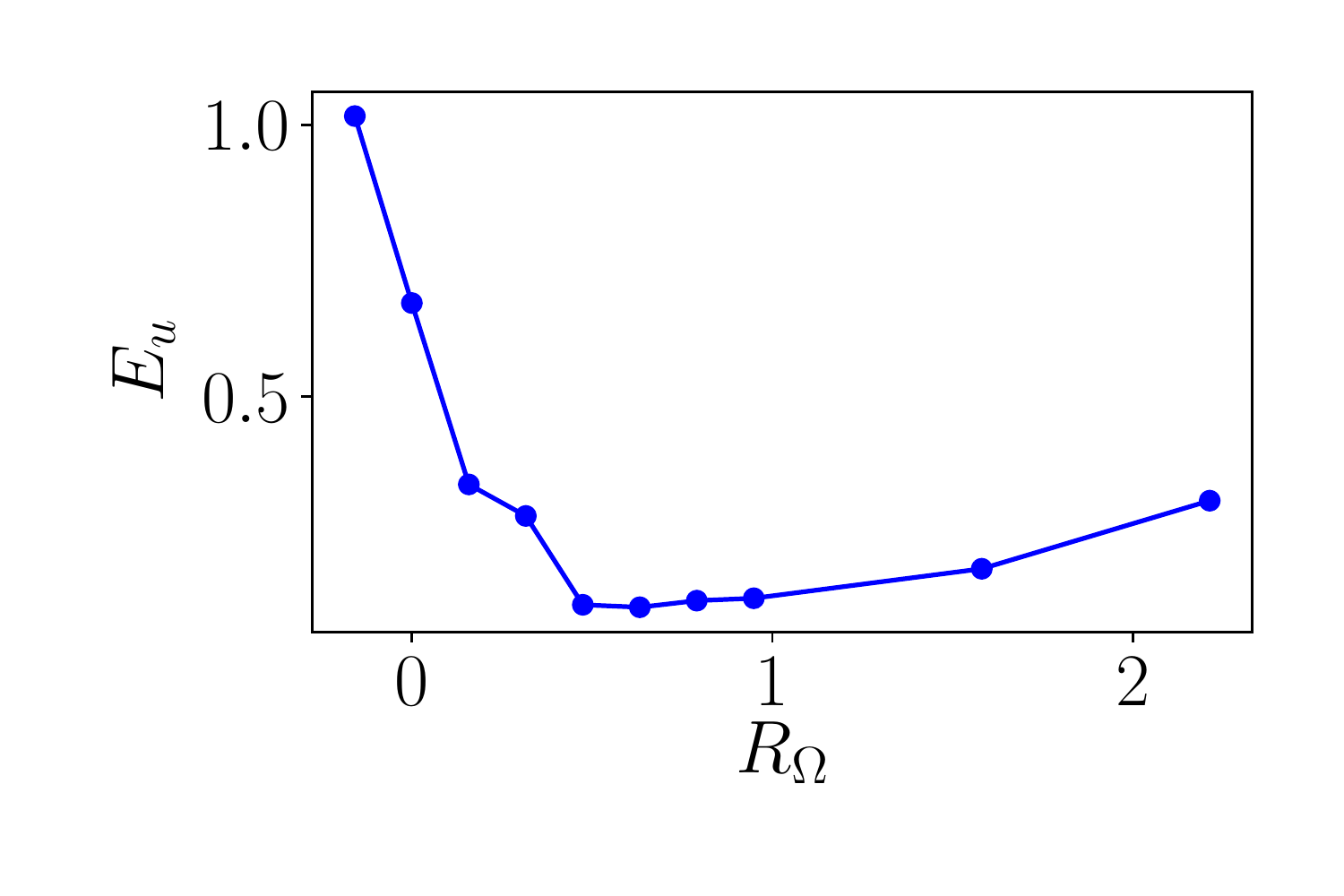}%
 \caption{ Left: Averaged streamwise velocity $\langle u \rangle$ for different values of $R_\Omega$. Right: Average kinetic energy of the mean streamwise flow $E_u$ for different values of $R_\Omega$. }
 \label{fi:AvgVel} 
 \centering
\end{figure}

We now analyze the effect of rotation on the velocity profiles. In the left panel of figure \ref{fi:AvgVel}, we show the averaged streamwise velocity profiles between the free-slip walls for various rotation numbers $R_\Omega$. First, it can be seen that the velocity profiles are symmetric around the mid-gap ($y/d=0.5$), and that the slip velocity at the walls is maximum for $R_\Omega= -0.16$. As the rotation number increases, the velocity profiles show a decrease in the slip-velocity until a minimum is reached at around $R_\Omega= 0.63$.

To quantify this, we define $E_u$ the average streamwise kinetic energy of the flow as:

\begin{equation}
 E_u = \frac{1}{2d} \displaystyle\int_0^d  \langle u \rangle^2  dy = \displaystyle\frac{1}{2} U^{*2},
 \label{eq:ustar}
\end{equation}

\noindent with $U^*$ a characteristic streamwise mean velocity. This $U^*$ is a response of the system. The dependence of $E_u$ on $R_\Omega$ is shown in figure \ref{fi:AvgVel}. As anti-cyclonic rotation is introduced, a prominent decrease of $E_u$ is seen, with a flat plateau around $R_\Omega=0.4-1.5$, after which $E_u$ increases again. This means that the same amount of momentum $T_{uv}$ (response) is transferred with a smaller velocity (input). We can use this to define the optimum shear transport as the value of $R_\Omega$ for which $E_u$ is minimum. With this interpretation, we can say that for our simulations of rotating Waleffe flow,  optimal transport appears as a broad ``peak'' (which is actually a minimum) in a large range of $R_\Omega$, similar to what was observed in low-Reynolds number plane Couette flow by \cite{Brauckmann16}.  The minimum $E_u$ is located at around $R_\Omega\approx 0.63$, but that minimum lies on a relatively smooth valley in the range $R_\Omega \in (0.4,1.0)$.

\begin{figure}
\includegraphics[trim=0cm 0 0cm 0, clip,width=0.4\textwidth]{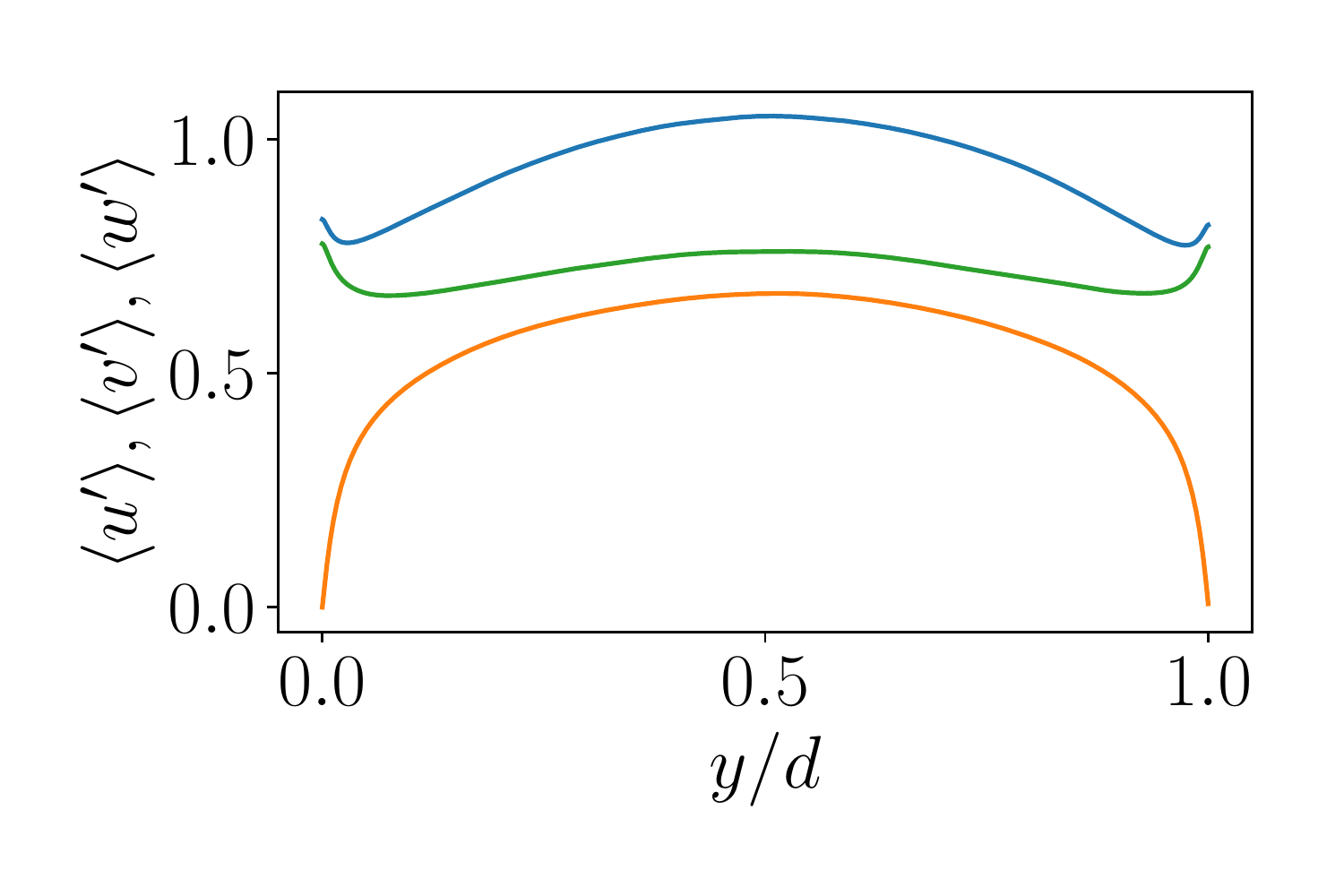}%
\includegraphics[trim=0cm 0 0cm 0, clip,width=0.4\textwidth]{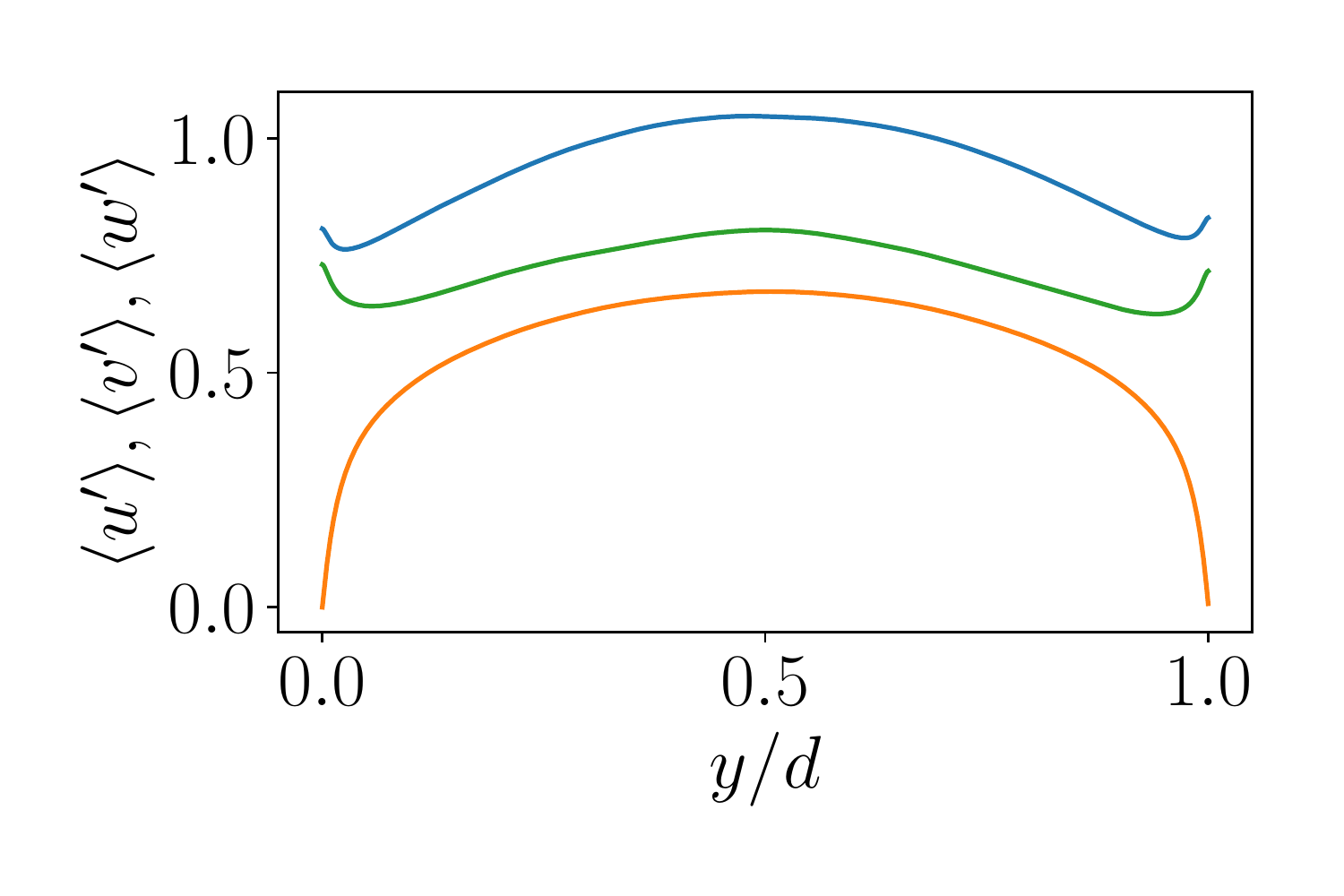} \\
\includegraphics[trim=0cm 0 0cm 0, clip,width=0.4\textwidth]{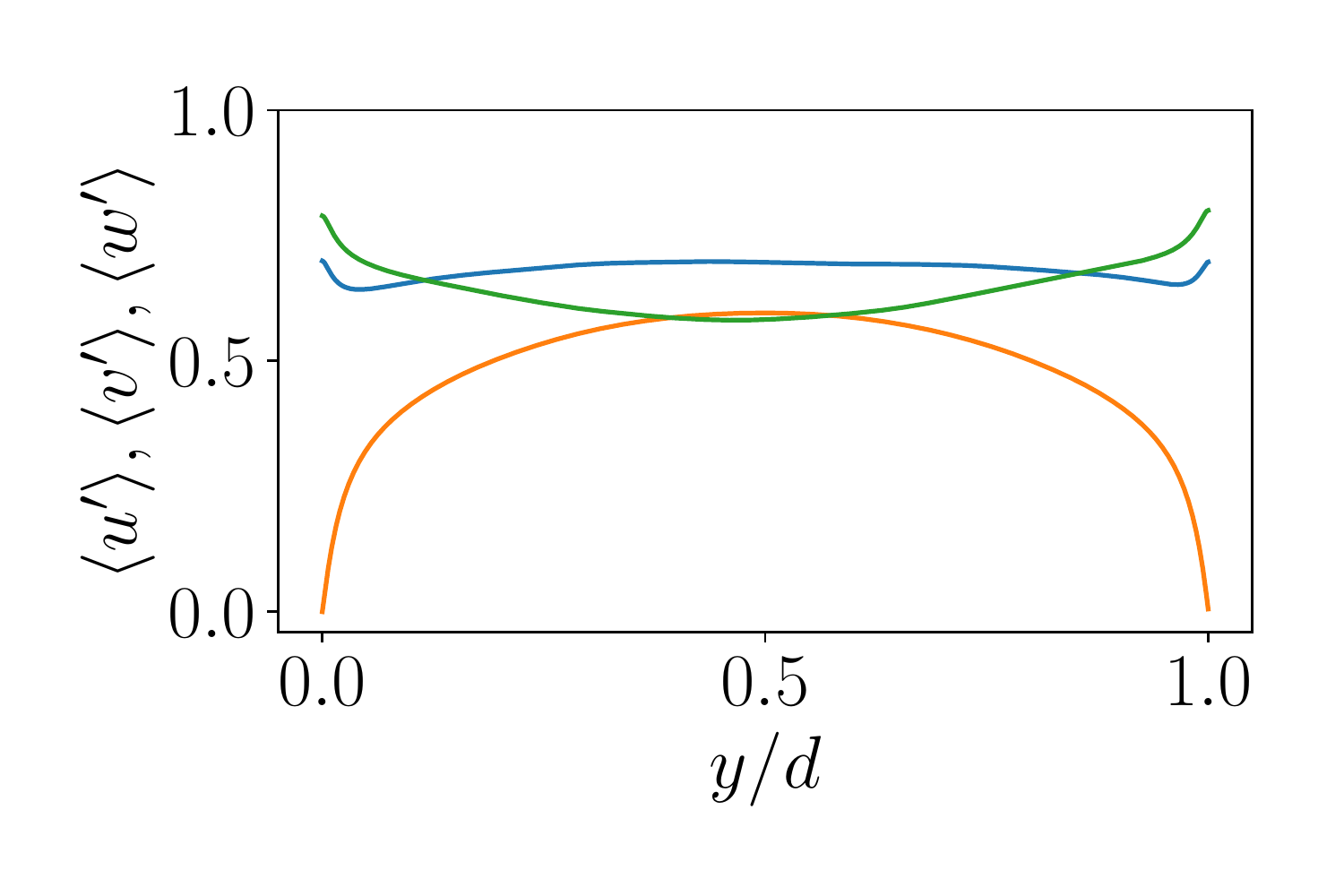}
\includegraphics[trim=0cm 0 0cm 0,
clip,width=0.4\textwidth]{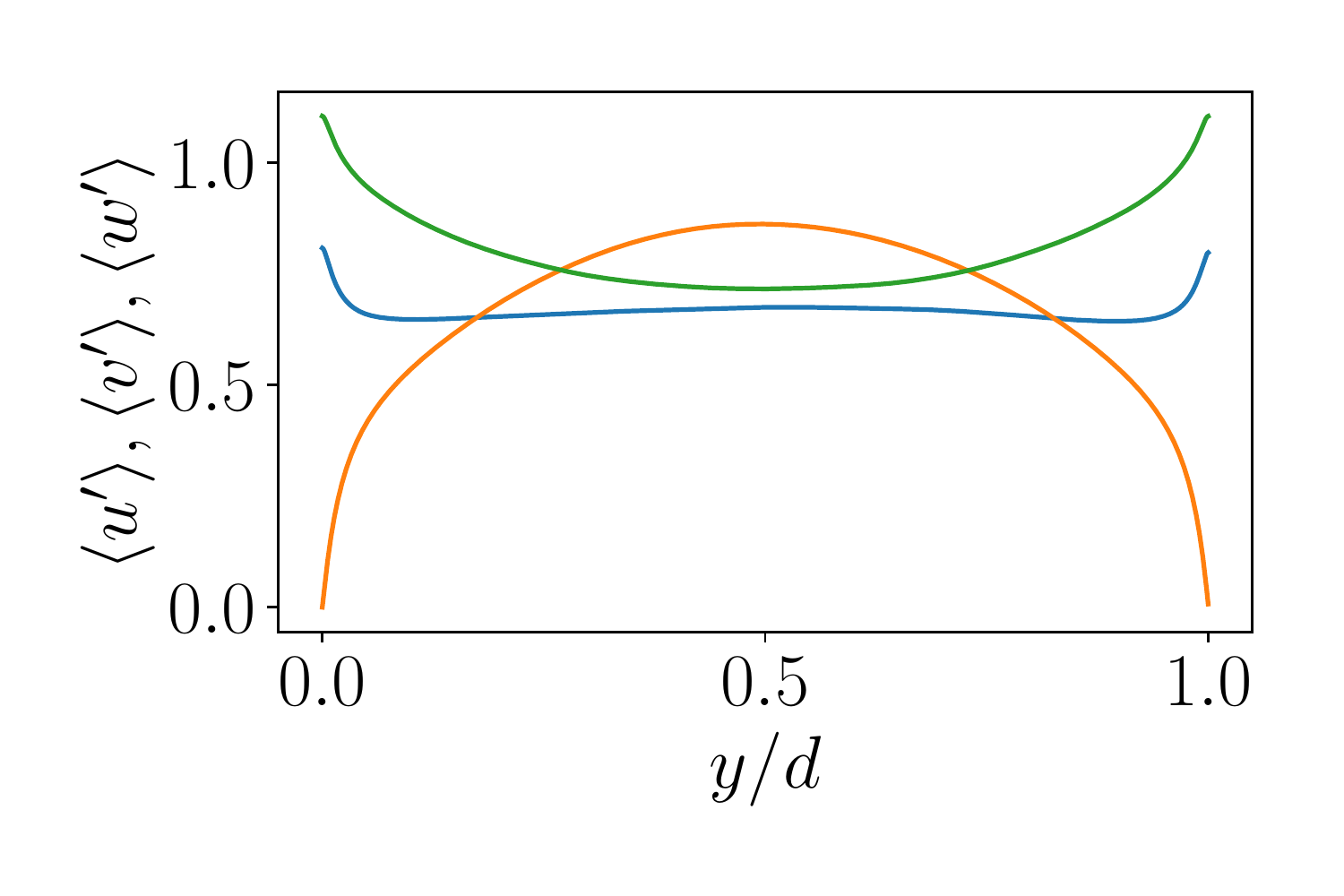}
\includegraphics[trim=13cm 105 0cm 0,
clip,width=0.085\textwidth]{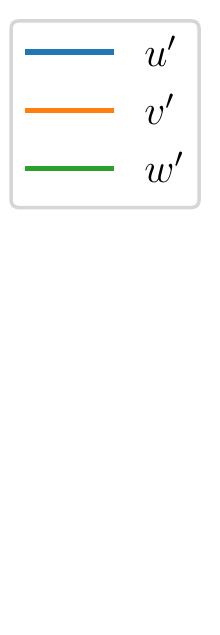}\\
\centering
\caption{ Root mean square velocity fluctuations at $R_\Omega =-0.16$ (top-left), $R_\Omega =0$ (top-right), $R_\Omega =0.63$ (bottom-left), and $R_\Omega =2.21$ (bottom-right).}
\label{fi:RMSVel} 
\end{figure}

To understand the mechanisms behind the optimal transport, we turn to the velocity root-mean-square velocity fluctuation profiles, shown in Figure \ref{fi:RMSVel} for some selected values of $R_\Omega$. The first thing we notice is the absence of a near-wall peak in the streamwise velocity fluctuation profiles. Instead a local minimum is seen in some, but not all cases. This suggests the absence of ordinary boundary layers due to the stress-free boundary coundition. We expect that the high Reynolds number boundary layer instability behind the narrow peak optimum transport in plane Couette flow \citep{Brauckmann17} will be absent. 

The second thing we notice is that cyclonic and no rotation, the streamwise velocity fluctuations $\langle u^\prime \rangle$ are largest of the three components at the mid-gap, but as $R_\Omega$ increases, the largest fluctuations become the ones in the wall-normal direction ($\langle v^\prime \rangle$). The largest streamwise velocity fluctuations $\langle u^\prime \rangle$ appear when $R_\Omega=-0.16$, i.e.~$R_\Omega$ is minimum. As $R_\Omega$ is increased, the values of $\langle u^\prime \rangle$ monotonically decrease. The smallest value of fluctuations corresponds to the largest value of $R_\Omega=2.21$. However, for the wall-normal velocity fluctuations the opposite pattern is seen, and the values of $\langle v^\prime \rangle$ increase with increasing $R_\Omega$ up to $R_\Omega=2.21$. The spanwise fluctuations show no discernible pattern in their variation with $R_\Omega$. This gives a hint to the mechanism behind optimum transport: the Coriolis forces due to spanwise rotation appear with different signs in the streamwise ($x$) and wall-normal ($y$) components of the Navier-Stokes equations, in one case increasing the fluctuations, in the other decreasing them.

\subsection{The effect of rotation on large-scale structures}
\label{sec:rotlsc}

\begin{figure}
\includegraphics[width=0.48\textwidth]{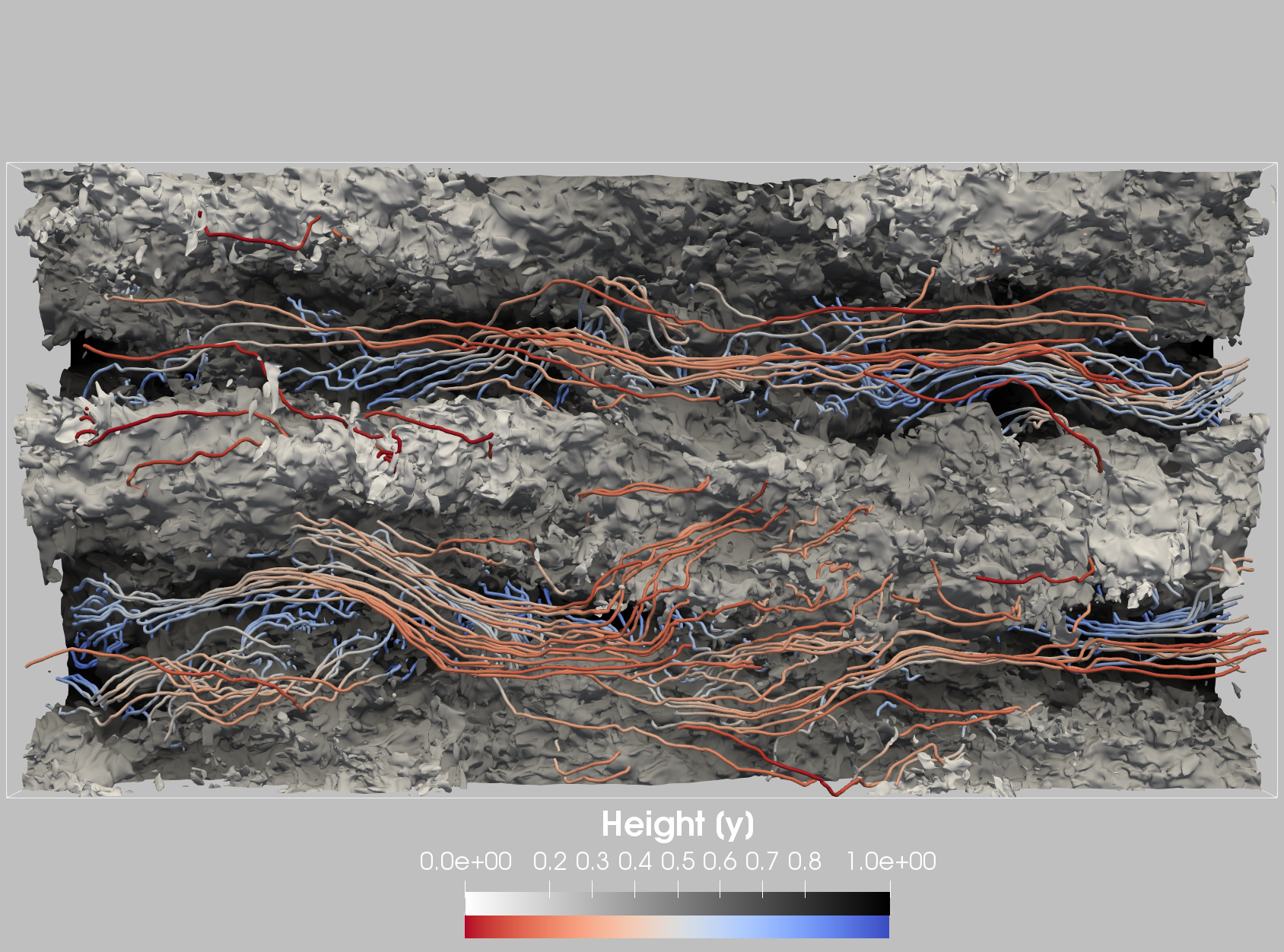}
\includegraphics[width=0.48\textwidth]{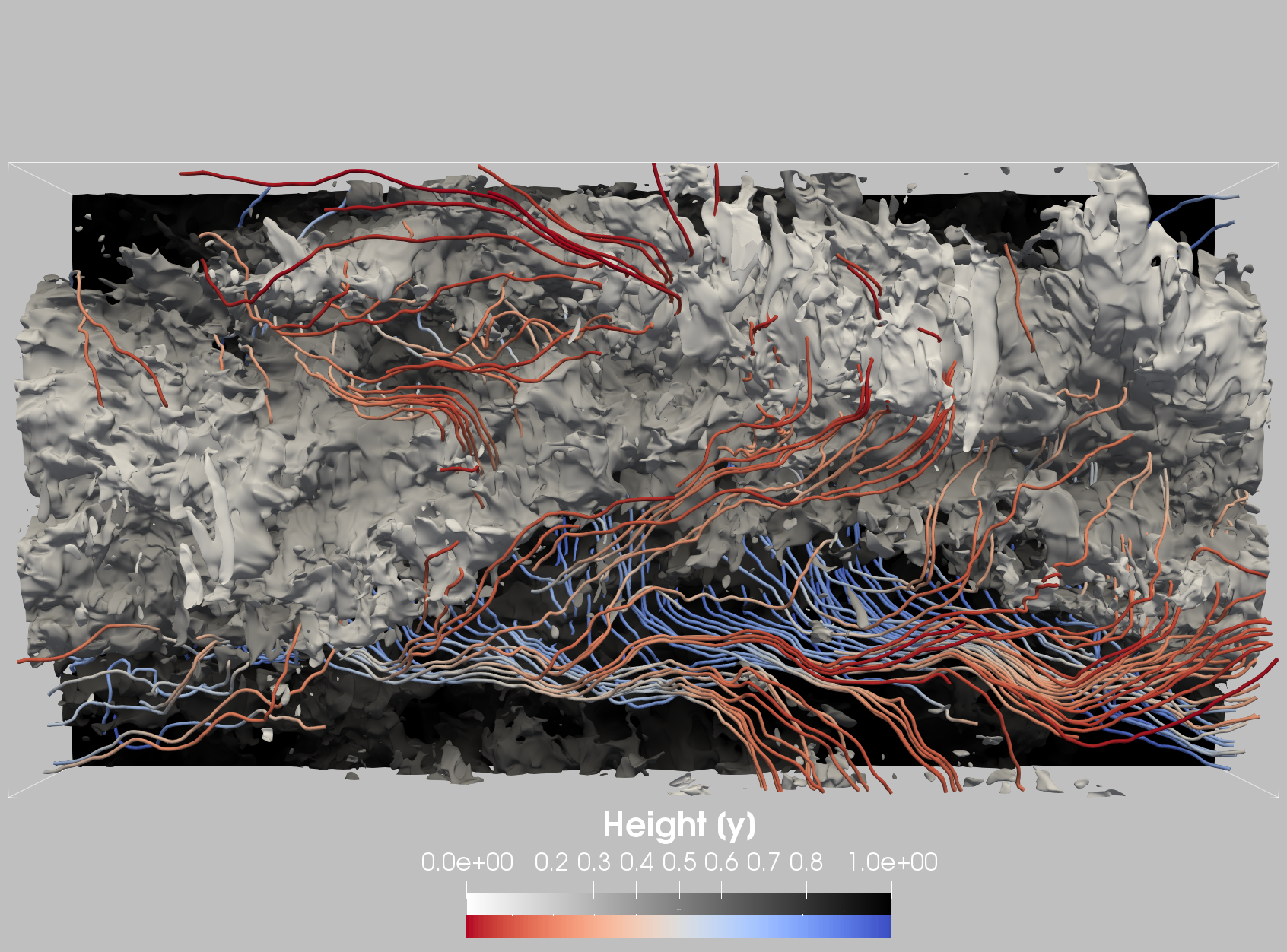}
\caption{Contour of instantaneous streamwise velocity $u$ at $R_\Omega=0.0$ (left, contour at $u=0.15$) and $R_\Omega=0.63$ (right, contour at $u=-0.1$). The view is projected onto a $x$-$z$ plane, with the $y$ coordinate represented through color. Flow is from left to right. The instantaneous streamlines are added, and colored according to $y$-coordinate. }
\label{fi:instant_U} 
\end{figure}

We now turn to the effect of rotation on large-scale structures. A visualization of the instantaneous streamwise velocity is shown in Figure \ref{fi:instant_U}, with instantaneous streamlines superimposed. We first observe the presence of large-scale structures in the flow for both the rotating and non-rotating cases. This could be expected from the autocorrelations in Figure \ref{Aut:Waleffe}. 

Large-scale flows can be considered secondary flows if their velocity components are perpendicular to the main flow direction. This is not always the case. In the absence of rotation, the streamwise velocity contour in figure \ref{fi:instant_U} is largely invariant in the streamwise direction. The secondary flow, i.e.~the cross flow in the wall normal and spanwise directions, is very weak. This can be deduced from  the relatively straight path of the streamlines. From this, we do not expect it to play a role in transporting shear.

As anti-cyclonic rotation is introduced, the flow is heavily modified. The streamwise velocity contour now has a completely different shape. More importantly, the secondary flow is strengthened, as can be seen from the visualized streamlines which move more in the wall-normal and spanwise direction. This secondary flow is of crucial importance as it helps with the transport of shear.  

To analyze these structures more quantitatively, we proceed in an analogous manner to \cite{Fran19}. Secondary flows are captured through the streamwise vorticity, as they are perpendicular to the main flow direction. By looking at the streamwise and temporal average of this quantity ($\Omega_x$) we capture only the spanwise-pinned streamwise-invariant structures, which are known to be the most relevant for shear transport. If large-scale structures are moving around the computational domain, they would not be captured by $\Omega_x$ as the averages would vanish.

\begin{figure}
\includegraphics[trim=7cm 15 0cm 0, clip, height=0.3\textwidth]{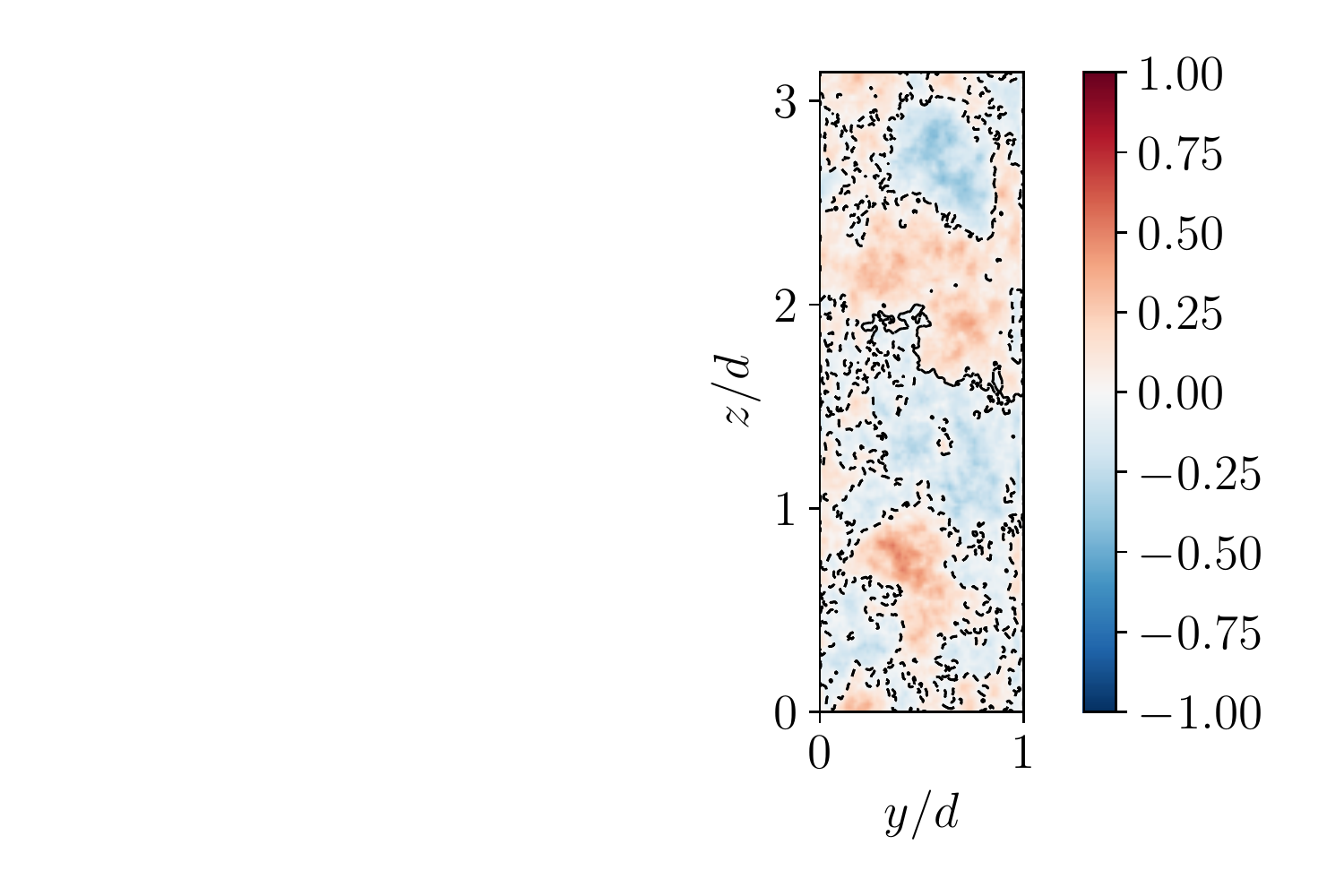}%
\includegraphics[trim=7cm 15 0cm 0, clip, height=0.3\textwidth]{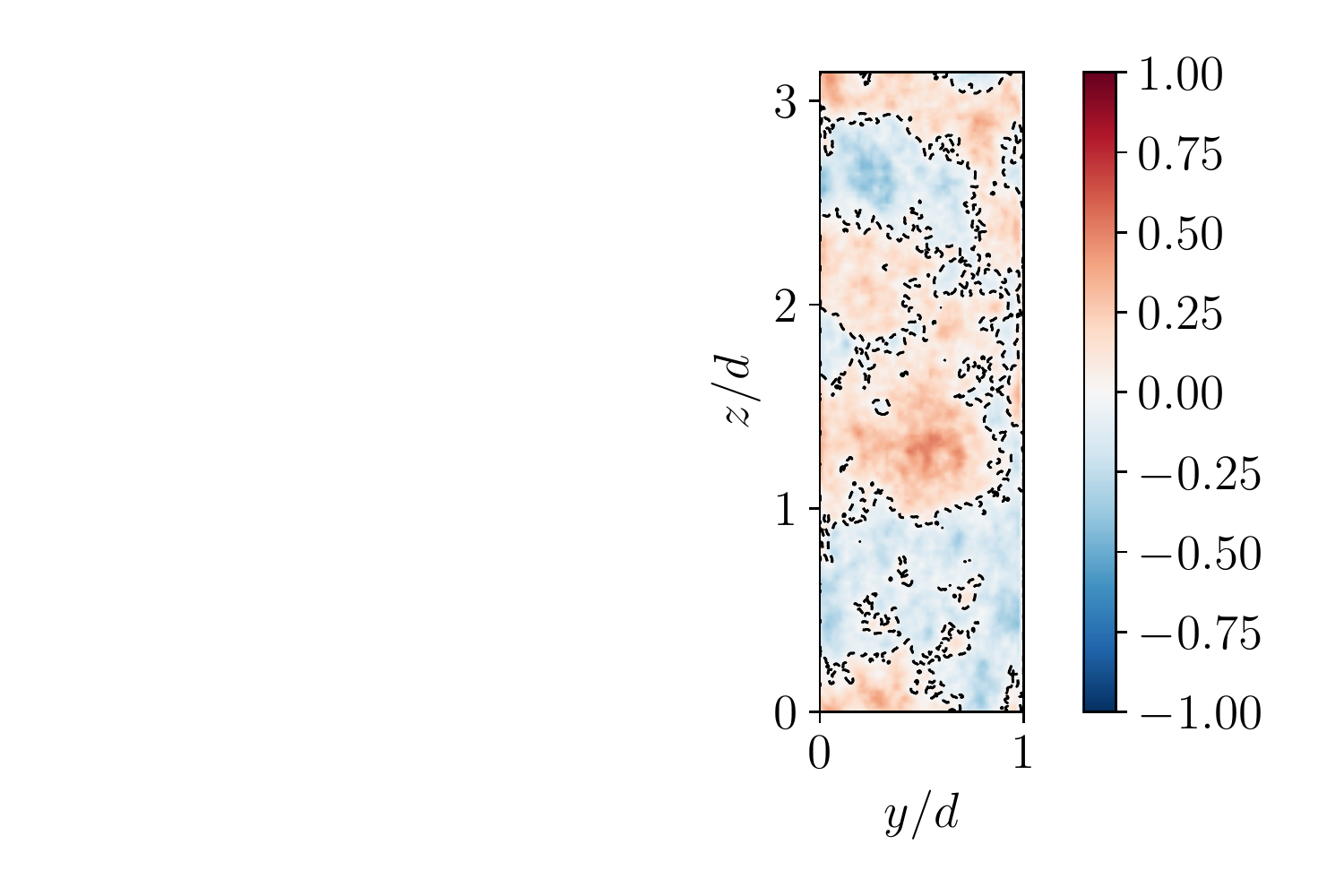}%
\includegraphics[trim=7cm 15 0cm 0, clip, height=0.3\textwidth]{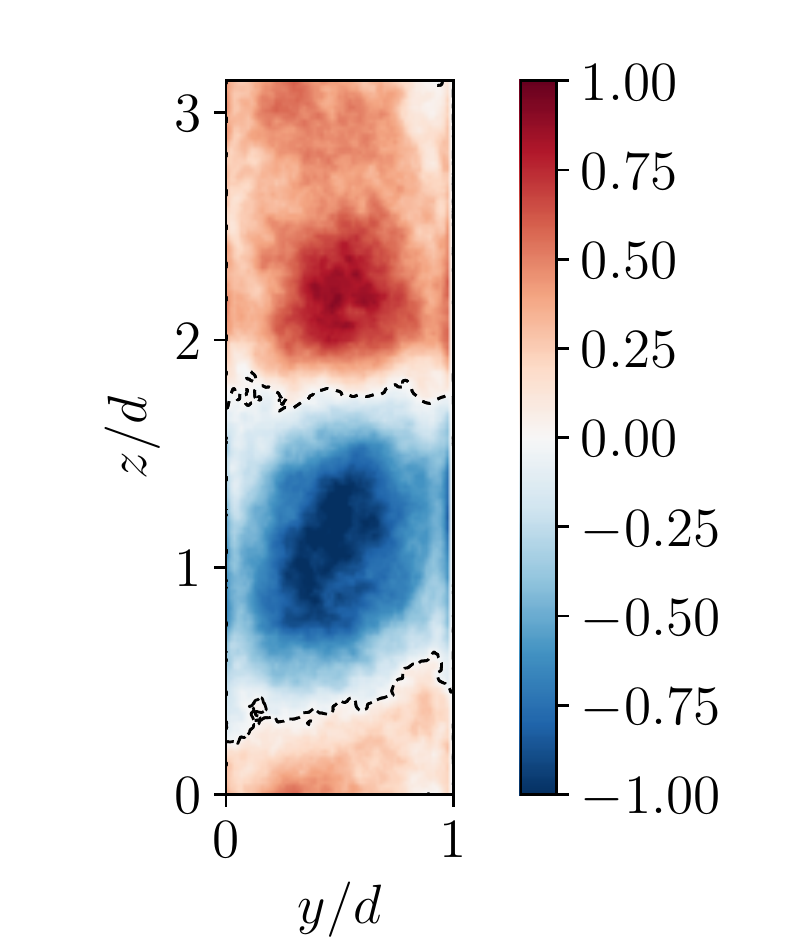}%
\includegraphics[trim=7cm 15 0cm 0, clip, height=0.3\textwidth]{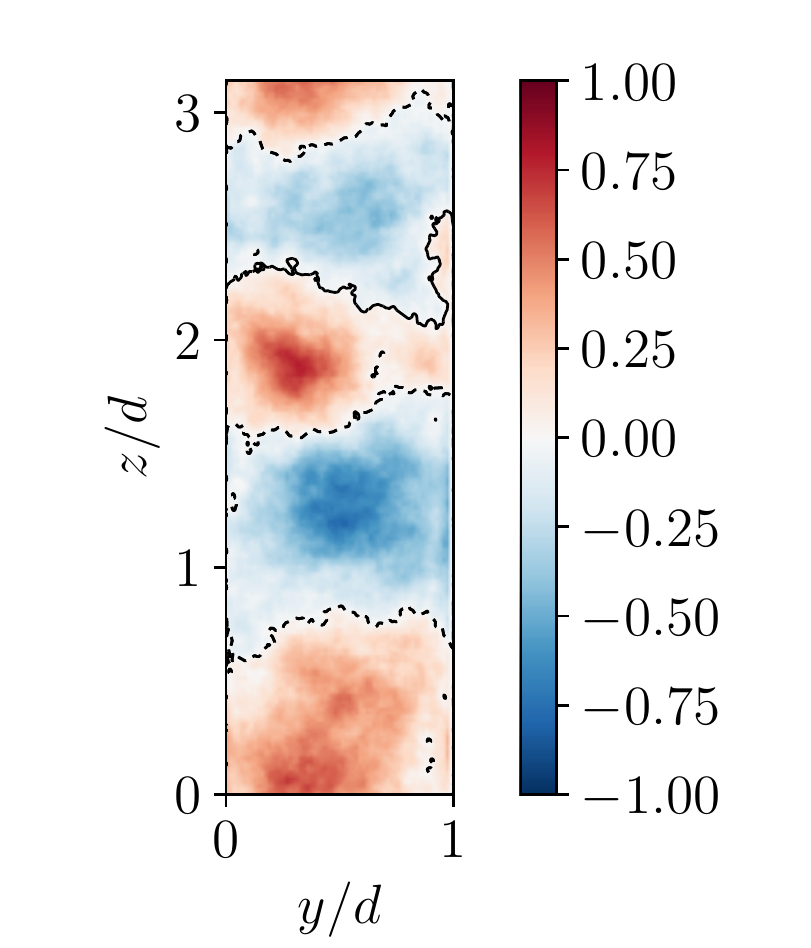}%
\\
\includegraphics[trim=0cm 0 0cm 0, clip, width=0.5\textwidth]{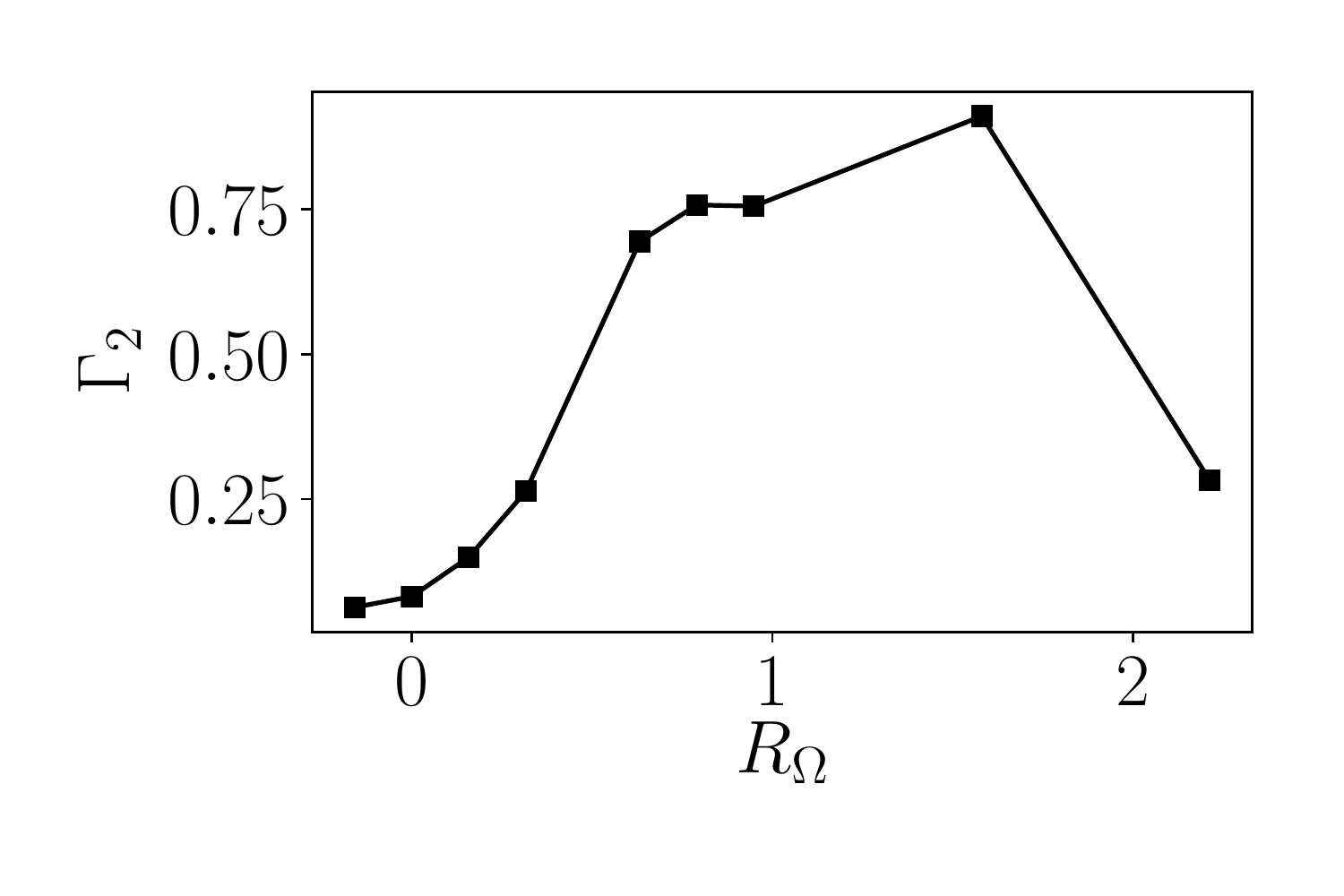}
\includegraphics[trim=0cm 0 0cm 0, clip, width=0.5\textwidth]{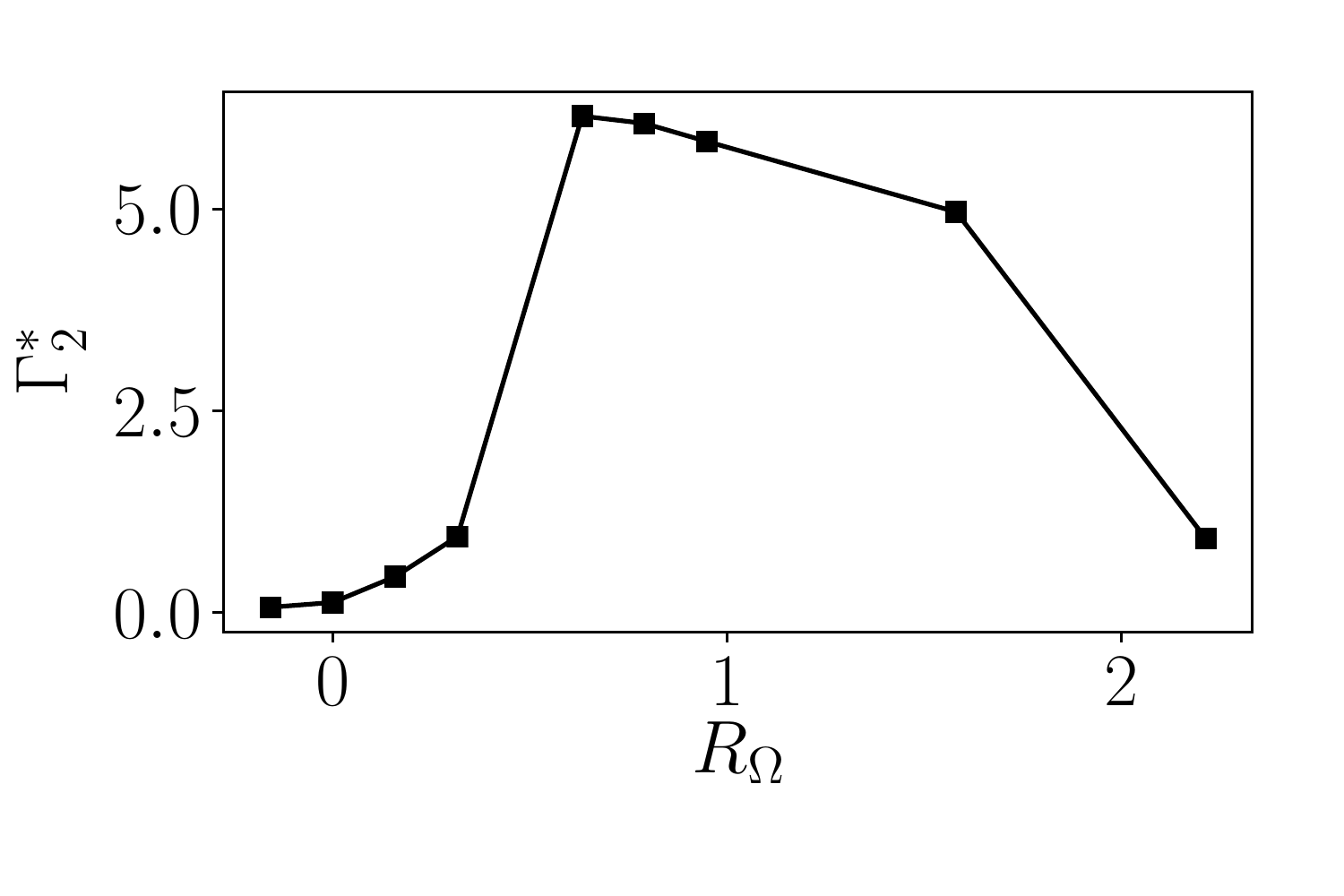}
\caption{The panels in the top row represents the streamwise and temporally averaged streamwise vorticity $\Omega_x $ for $R_\Omega=-0.16$, $0.0$ $0.63$ and $2.21$ (left to right). A black contour at zero has been added to delineate negative vorticity from positive vorticity regions. The left bottom panel shows the averaged circulation energy $\Gamma_2$ for thelarge-scale structures in the range of $R_\Omega$ $\in$ [-0.16, 2.21], and the bottom right panel depicts the effective averaged circulation energy $\Gamma_2^*$ in the same $R_\Omega$ range.}
\label{fi:Energy} 
\end{figure}

Figure \ref{fi:Energy} shows $\Omega_{x}$ for different values of $R_\Omega$. As hinted by Figure \ref{fi:instant_U}, spanwise-pinned and streamwise invariant secondary flows, with a vorticity core, appear as anti-cyclonic rotation is introduced. The strength of the roll-like structures appears to increase with increasing anti-cyclonic rotation. At around  $R_\Omega \approx 1.5$, the trend changes, and further increasing the rotation makes the structures unorganized, as shown in the right most panel at $R_\Omega=2.21$.

To quantify the strength of the rolls, we define the mean-squared circulation of the rolls as the volume integral:

\begin{equation}
\Gamma_2 = \frac{1}{d~L_z} \displaystyle\int_0^{d} \int_0^{L_z} \Omega_x^2 ~dy~dz.
\end{equation}

\noindent and show this as a function of $R_\Omega$ in the bottom left panel of Figure \ref{fi:Energy}. We  can see how the strength of the pinned rolls increases with increasing $R_\Omega$ until the maximum is reached. However, using $R_\Omega$ to compare roll strength is not totally fair. As shown in section $\S$3.2, the underlying magnitudes of velocity change. Therefore, we define $\Gamma_2^*=\Gamma_2 \tilde{U}^2/U^{*2}$, where $U^{*}$, a characteristic streamwise velocity, was defined in equation \ref{eq:ustar}. We show $\Gamma_2^*$ as a function of $R_\Omega$ in the bottom right panel of Figure \ref{fi:Energy}, and observe that the strongest, and most ordered structures correspond to the range $R_\Omega=0.47-1$. This coincides with the range of $R_\Omega$ for which optimal shear transport takes place, and highlights the link between optimal transport and strongest large-scale structures as was seen in plane Couette flow by \cite{Brauckmann13,Brauckmann17}.

\subsection{Large-scale structures and the self-sustained process}
\label{sec:tevolsc}

\begin{figure}
\includegraphics[trim=7cm 19 0cm 0, clip, height=0.5\textwidth]{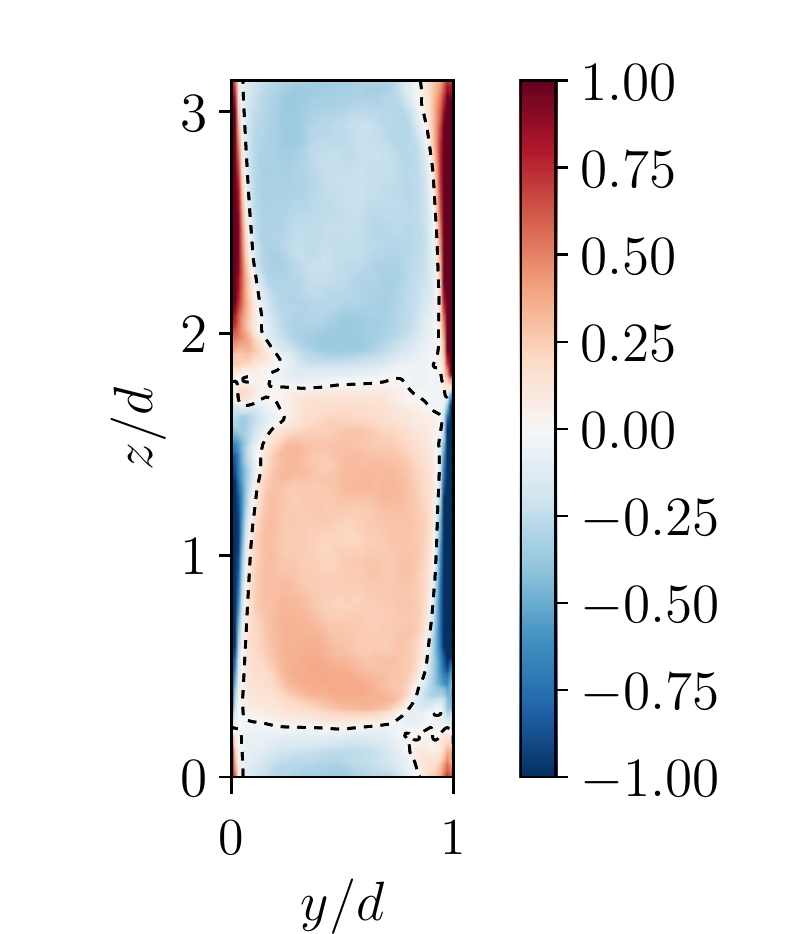}%
\includegraphics[trim=7cm 19 0cm 0, clip, height=0.5\textwidth]{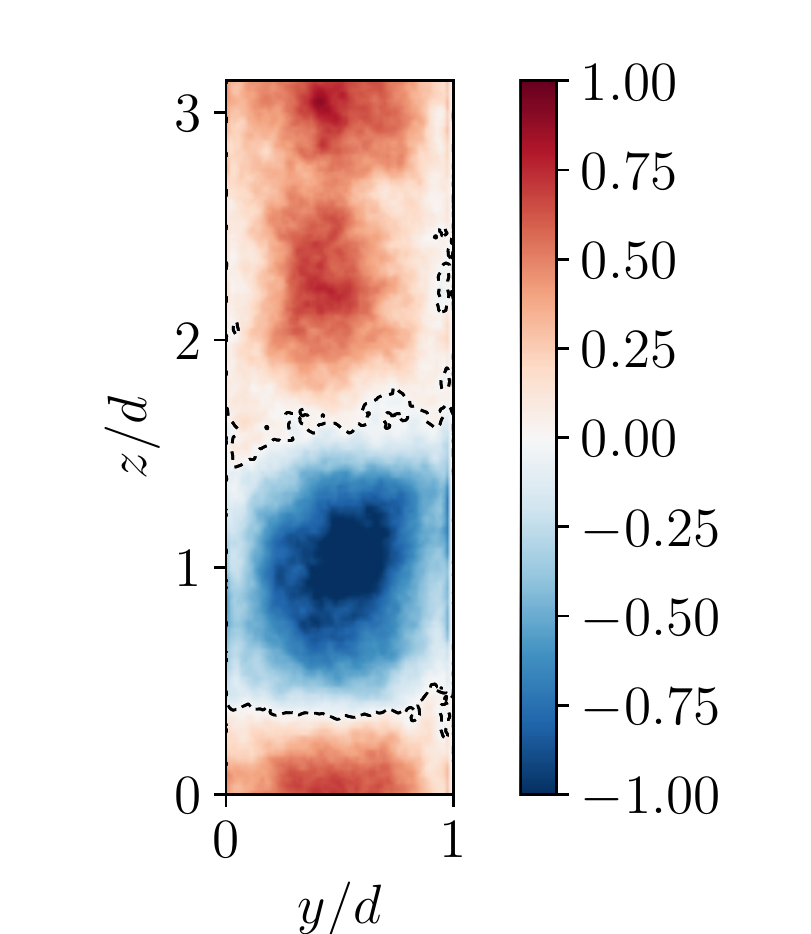}%
\caption{Pseudocolor plot of temporally averaged streamwise vorticity $\Omega_x$  for plane Couette flow at $R_\Omega=0.1$ (left) and rotating Waleffe flow at $R_\Omega=0.63$ (right). Contours levels for vorticity are shown at zero to highlight the large-scale structures.}
\label{fi:Vorticity} 
\end{figure}

First, we visually compare the rolls in rotating Waleffe flow to those in rotating plane Couette flow in Figure \ref{fi:Vorticity} using $\Omega_x$. One thing we notice is these structures are more clearly defined in plane Couette flow than Waleffe flow. Furthermore, the streamwise vorticity of the rolls is localized in the core of the structures in Waleffe flow. The streamwise vorticity in plane Couette flow is localized in mainly in the boundaries of the structure, with additional vorticity being generated by the boundary layers. Nevertheless, these results show that pinned large-scale structures arise only with a generic shear and anti-cyclonic rotation. 

We can probe the temporal behaviour of the large-scale structure. In \cite{Fran19}, a link between the large-scale fixed structures in Taylor-Couette, i.e. the turbulent Taylor rolls, and the self-sustained process was found by analyzing the rolls in Fourier space following \cite{ham95}. While the analogy was far from perfect, the mean flow energy was found to oscillate in anti-phase with the spanwise fundamental mode, which represented the large-scale structure through its rolls and streaks. 

We proceed in the same manner, and we begin by defining the modal r.m.s.~velocity as:

\begin{equation}
    M\left(k_x=\alpha m,k_z=\beta n\right)=\left[ \displaystyle\int_{0}^{d} \left( \hat{u}^2(m \alpha,y,n \beta) +\widehat{v}^2(m \alpha,y,n \beta)+ \hat{w}^2(m \alpha,y,n \beta) \right)  dy \right]^{\frac{1}{2}},
\end{equation}

\noindent where $\alpha=2 \pi/L_x$ and $\beta=2\pi/L_z$ are the fundamental streamwise and spanwise wavenumbers, and $\hat{\phi}$ represents the Fourier transform of $\phi$ in spanwise and streamwise directions. We focus on two modes: $M(0,0)$, the spanwise and streamwise invariant mode, which represents the mean flow, and $M(0,\beta)$ the streamwise independent, fundamental in $z$, corresponding to the large-scale structure.

\begin{figure}
\includegraphics[width=0.45\textwidth]{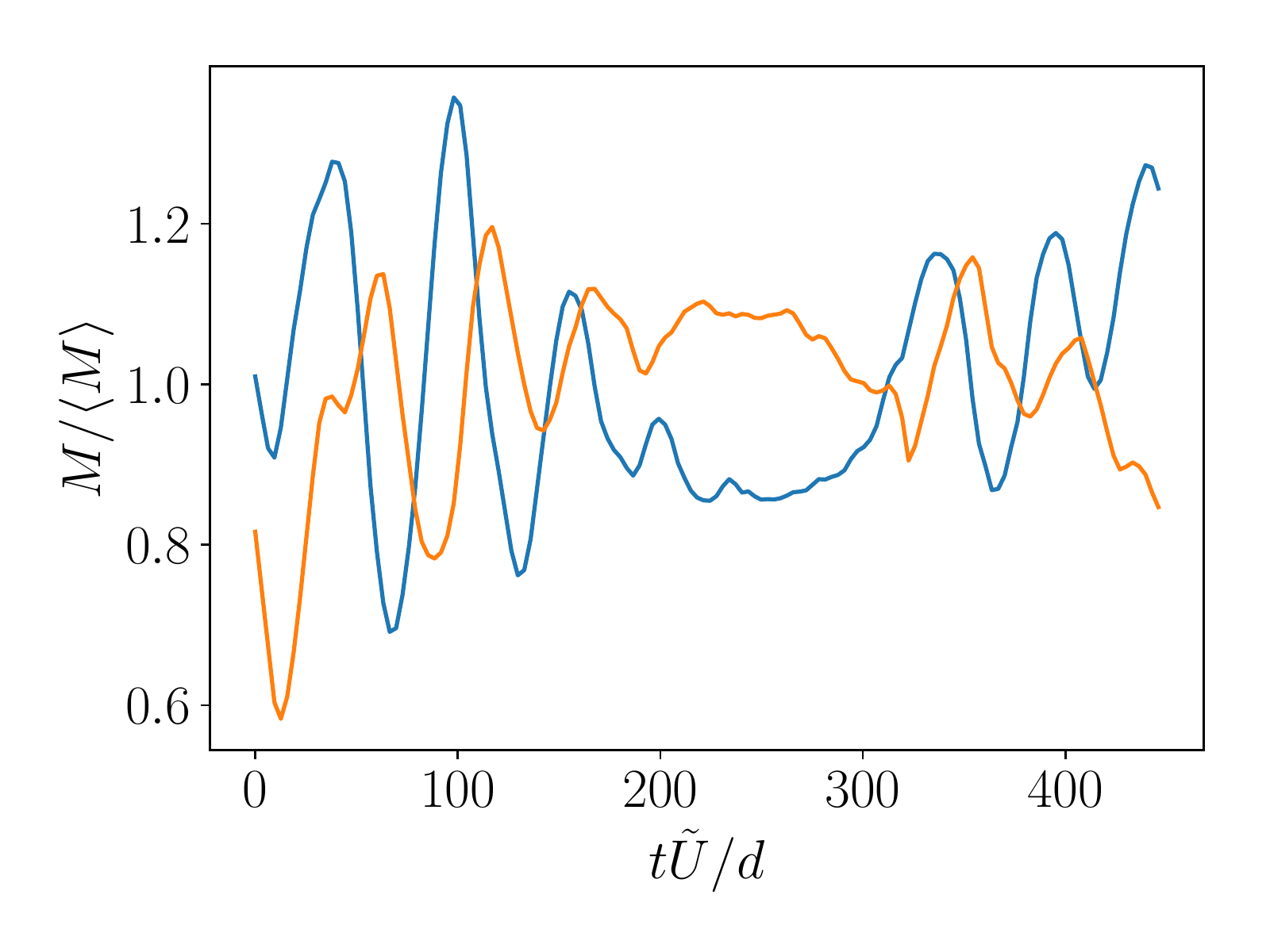}
\centering
\caption{Temporal evolution of the normalized modal RMS velocity in the Fourier space associated to the $M(0,\beta)$ mode (orange) and the $M(0,0)$ mode (blue) for $R_\Omega=0.63$.} 
\label{fi:tevo_streak_roll} 
\end{figure}

In Figure \ref{fi:tevo_streak_roll} we see that the energies of these two modes oscillate at time-scales of $\mathcal{O}(50 d/\tilde{U})$, and the period of the two quantities is almost anti-correlated, consistent with the breakdown-regeneration structure of shear flows described in \cite{ham95}, and the behaviour of the turbulent Taylor rolls in \cite{Fran19}. Energy is constantly being redistributed from the mean flow into the streaks and rolls of the large-scale structure. Remarkably there are some ``dead'' times ($t\tilde{U}/d=200-300$) where the cycle is temporarily broken and there is no significant exchange of energy.

We can probe the origin of the structures following the procedure set out in \cite{pir18}. Secondary flows are perpendicular to the main flow, and as such will have vorticity mainly in the main flow direction, i.e.~ the streamwise direction. Starting off with a Reynolds-Averaged (in the streamwise and time coordinates) equation for $\Omega_x$  \citep{einstein1958}, 

\begin{equation}
    \overline{v} \displaystyle\frac{\partial \Omega_x }{\partial y} +  \overline{w} \displaystyle\frac{\partial \Omega_x }{\partial z} = \left (  \displaystyle\frac{\partial^2 }{\partial y^2} -  \displaystyle\frac{\partial^2 }{\partial z^2} \right)(-\langle \overline{v'w'} \rangle) +  \displaystyle\frac{\partial^2 }{\partial y \partial z}( \overline{v^{\prime 2}}  - \overline{w^{\prime 2}}) + \nu \left (  \displaystyle\frac{\partial^2 \Omega_x }{\partial z^2} + \displaystyle\frac{\partial^2 \Omega_x  }{\partial z^2} \right ),
\end{equation}

\noindent where $\overline{\phi}$ denotes a temporal and streamwise average, $\phi^\prime$ are fluctuations around that mean, and $\overline{\omega}_x$ is simply $\Omega_x$.

The various terms in this equation are associated with the effect of mean cross-stream convection (left-hand side), secondary turbulent shear stress (first term on right-hand side), normal stress anisotropy (second term), and viscous diffusion (third term). If the convective terms vanish, the convective transport of average streamwise vorticity is zero. If this holds, it should be possible to write a streamfunction for the cross-flow secondary motions which has a strict functional relationship to the vorticity \citep{pir18}.

The streamfunction is evaluated by solving:

\begin{equation}
 \nabla^2\psi=-\Omega_x,
\end{equation}

\noindent with a constant Dirichlet boundary conditions at the walls, because the stress-free walls behave like a streamline. We take this free constant to be zero.

\begin{figure}
 \centering
\includegraphics[trim=7cm 0 0cm 0, clip, height=0.3\textwidth]{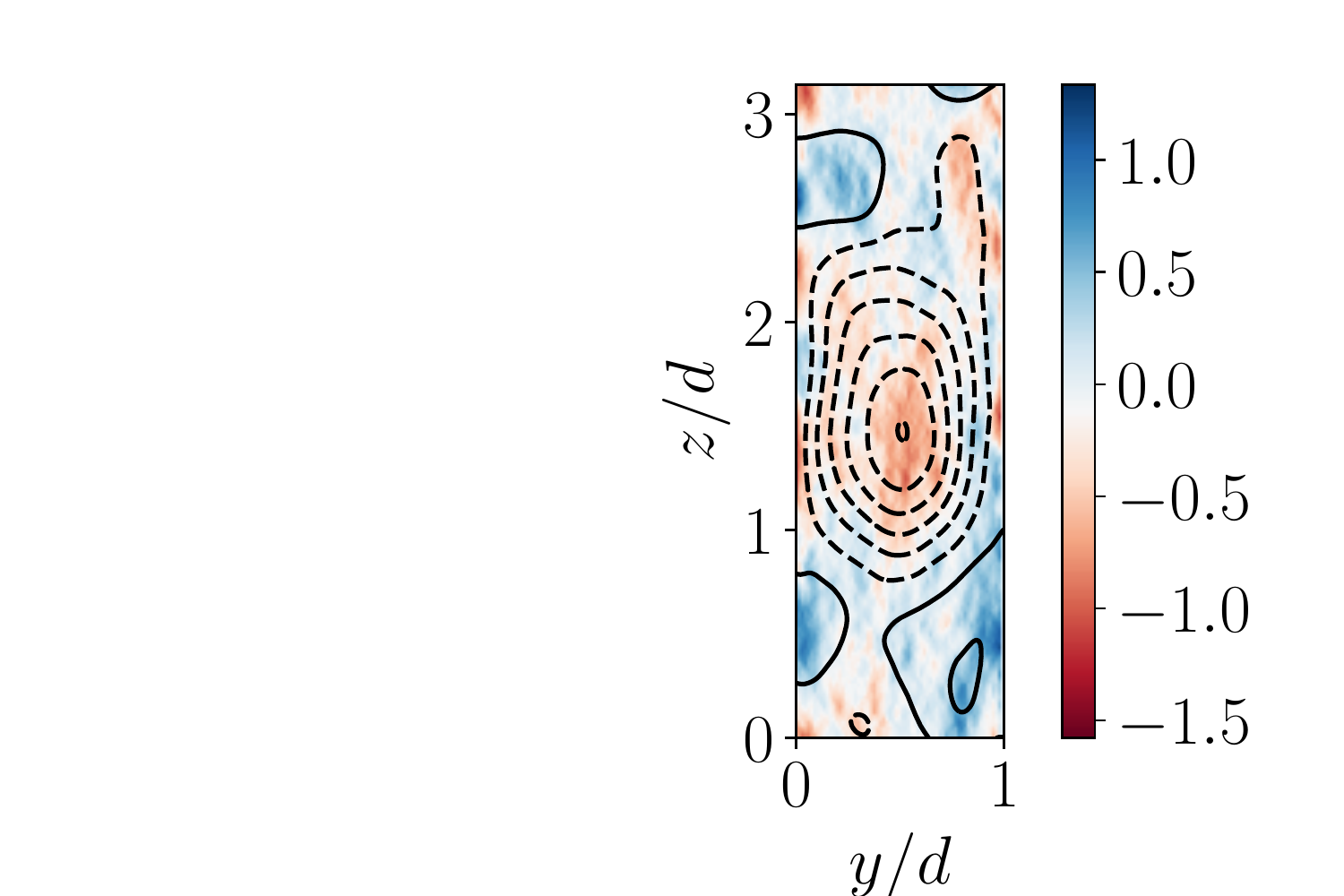}%
 \includegraphics[trim=0cm 0 0cm 0, clip, height=0.3\textwidth]{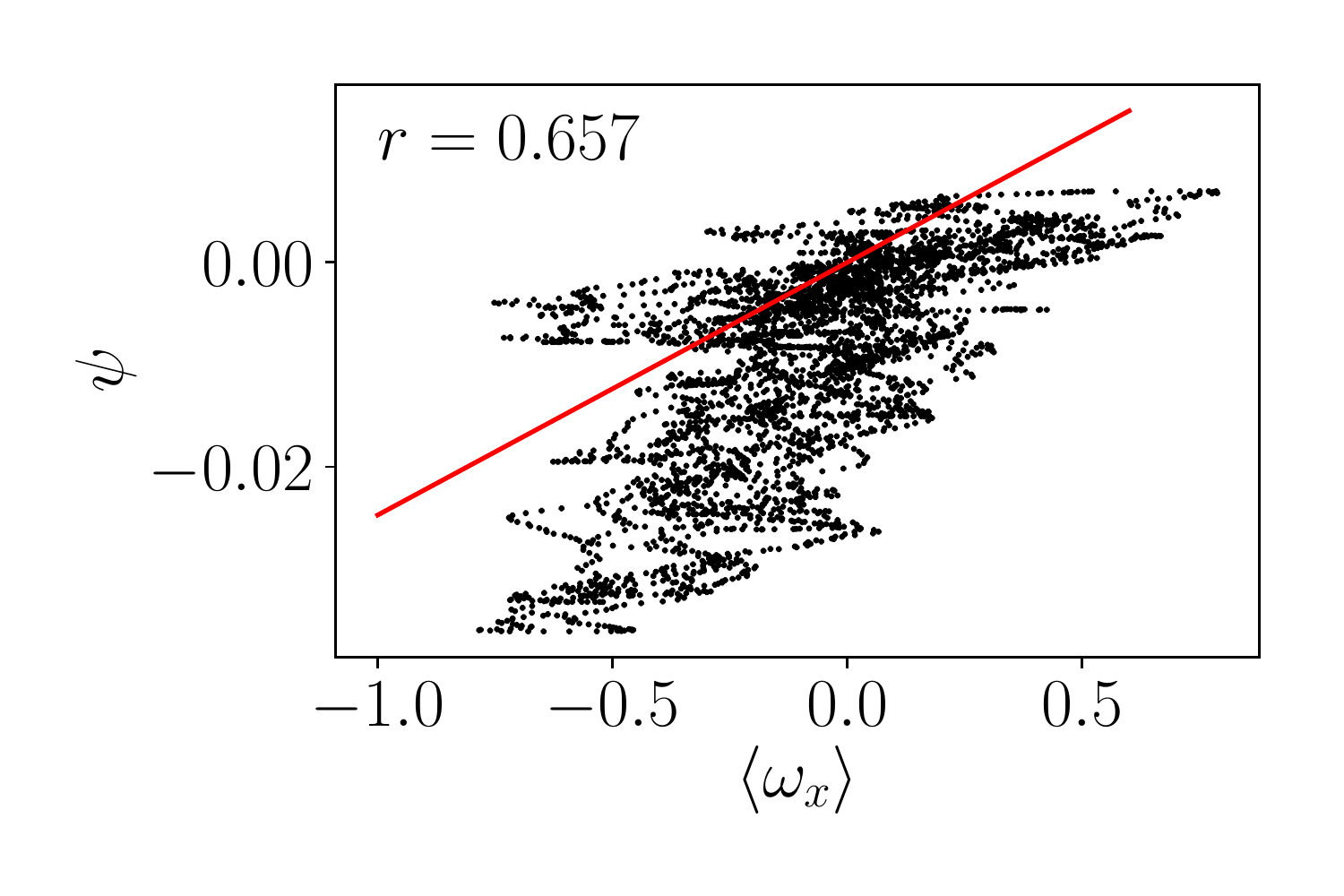}\\
 \includegraphics[trim=7cm 0 0cm 0, clip, height=0.3\textwidth]{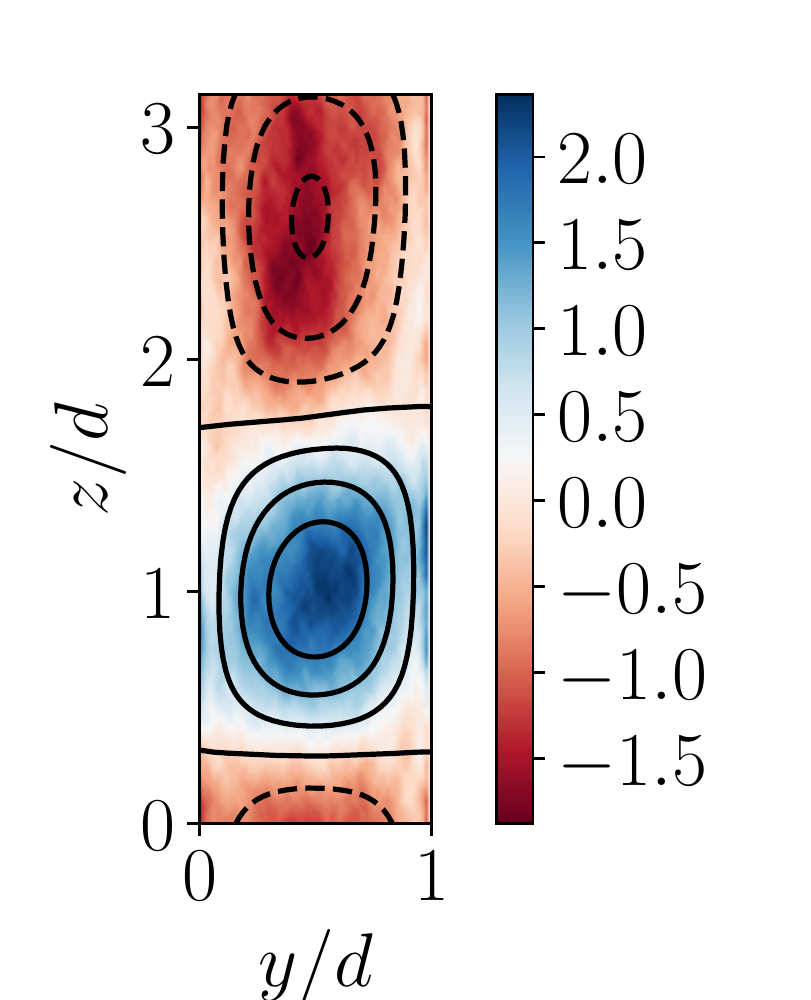}%
  \includegraphics[trim=0cm 0 0cm 0, clip, height=0.3\textwidth]{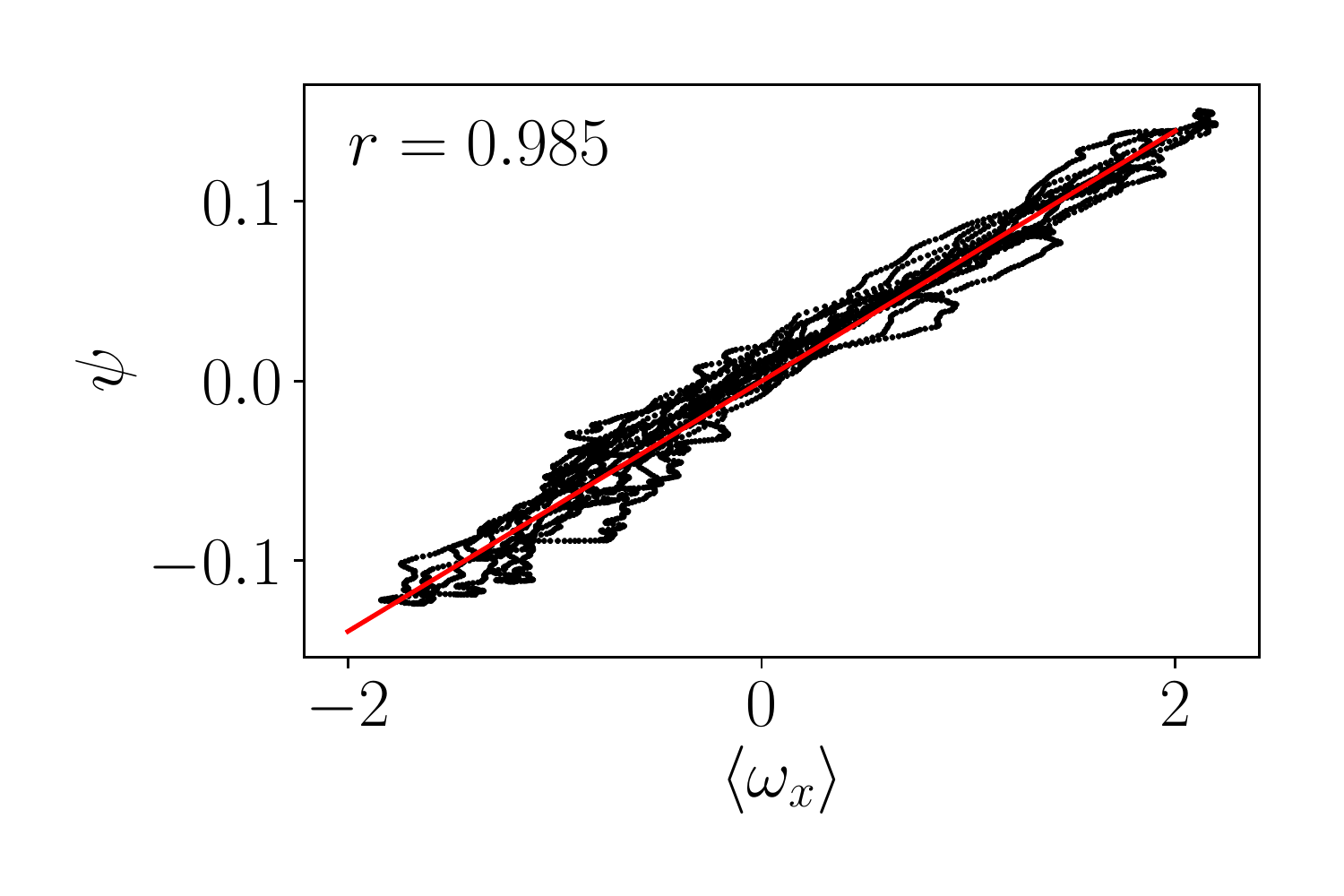}\\
 \caption{Left column: Pseudocolor plot showing $\Omega_x$ at $R_\Omega=0$ (top) and $R_\Omega=0.63$ (bottom) with contours of constant $\psi$ superimposed. The contours on the top plot are spaced $0.012$ units of $\psi$, while on the bottom plot they are spaced $0.04$ units of $\psi$, with dashed contours indicating negative values of $\psi$. Right column: Scatterplot of $\psi$ against $\Omega_x$ corresponding to the plots on the left. The best linear fit is plotted in red.}
 \label{fi:Streamfunction} 
\end{figure}

In the first two panels of figure \ref{fi:Streamfunction}, we superimpose contours of constant $\psi$ to a pseudocolor plot of $\Omega_x$. No clear relationship can be seen for $R_\Omega=0$, but the circular contours of $\psi$ at $R_\Omega=0.63$ overlap on to the large-scale structures of $\Omega_x$ making evident the relationship between $\psi$ and $\Omega_x$. The functional relationship between $\psi$ and $\Omega_x$ is shown more prominently in right bottom panel of figure \ref{fi:Streamfunction}, where a scatterplot of $\psi$ and $\Omega_x$ for $0.1<y/d<0.9$ is shown. A quasi-linear relationship between them can be seen in the regions far away from the wall. 

This can be understood following \cite{pir18}, who decomposed $\Omega_x$ as eigenfunctions of the Laplace operator:

\begin{equation}
 (\nabla^2 + k^2) \psi = 0.
\end{equation}

The admissible values of $k$ give us the different eigenfunctions of the Laplacian.  A linear regression, fit to data at $y/d\in[0.1,0.9]$, shows the best fit line coefficient ($k^2$) at $6.96 \times 10^{-2}$ giving the value of $\Omega_x = k^2 \psi$. This indicates that the fixed secondary motions in Waleffe flow correspond very well to a single eigenmode of the Laplacian operator. Overall, these results differ little from those obtained in plane Couette flow, showing that the behaviour of the large-scales is barely affected by the no-slip condition.

\section{Summary and conclusions}
\label{sec:conclusion}

We performed direct numerical simulation (DNS) of rotating Waleffe flow at a fixed $Re=3.61\times 10^3$. Once adequate resolution was determined, a study of the effect of domain size was performed. The decorrelation lengths in the spanwise and stream-wise directions were shown to have a strong dependence on the domain size. They further revealed that rotation does not substantially affect the velocity autocorrelations in the streamwise direction, but that it has a strong effect on the spanwise direction, as it modifies the large-scale structures.

Anti-cyclonic spanwise rotation modifies the shear transport, and an ``optimal'' shear transport appears at around $R_\Omega \approx 0.63$, in so much as the mean streamwise energy is reduced to a minimum for a fixed shear transport. This is because the transported shear has to equal the underlying forcing. Anticyclonic rotation also modifies the underlying velocity fluctuation profiles, heavily decreasing streamwise fluctuations and enhancing wall-normal fluctuations. 

Due to the absence of boundary layers, we linked optimal transport in rotating Waleffe flow to the broad peak of optimal shear transport in plane Couette flow found by \cite{Brauckmann16}. In a similar manner as discussed in \cite{Brauckmann13,Brauckmann16}, this ``peak'' is linked to the appearance and strengthening of pinned large-scale structures. Once the energy was corrected to account for the varying strength of the mean flow, these structures were found to be strongest at around the same values of $R_\Omega \in (0.4,1)$, which corresponds to the value of $R_\Omega$ that achieve optimal transport. The structures were found to periodically take energy from the mean flow to energize, and were also linked to eigenvalues of a streamfunction for secondary flows, following \cite{pir18}.

With these simulations we have shown that the appearance of streamwise invariant, spanwise pinned structures which increase transport are a generic characteristic of anticyclonic shear flows, appearing in both rotating Waleffe and rotating plane Couette. We note that these structures appear to be invariant in a direction \emph{normal} to rotation, unlike those which could be expected from the Taylor-Proudmann problem. An avenue for further research, aside from increasing the Reynolds number, is probing homogeneous shear turbulence to search for these structures. This system removes the last confinement on the structures, the no-penetration top and bottom walls.

\emph{Acknowledgments:} We thank B. Eckhardt, S. Pirozzolli, F. Sacco and R. Verzicco for fruitful discussions. We thank the Center for Advanced Computing and Data Science (CACDS) at the University of Houston for providing computing resources.

\bibliographystyle{jfm}
\bibliography{Waleffe}

\end{document}